\documentclass[letterpaper]{JHEP3}

\usepackage{amssymb}
\usepackage{amsmath}
\usepackage{graphicx}
\usepackage{bm} 
\usepackage{amssymb}
\usepackage{amstext}
\usepackage{tensor}
\usepackage{epsf,epsfig}
\usepackage{amsfonts} 
\usepackage{graphicx}
 \usepackage{bm} 
 \usepackage{cite}
\usepackage{amsmath}
\usepackage{mathrsfs}
 
\usepackage{amsmath,amssymb}

\newcommand{\insertplot}[5]{\begin{figure}
 \hfill\hbox to 0.05in{\vbox to #5in{\vfill
 \inputplot{#1}{#4}{#5}}\hfill}

 \hfill\vspace{-.1in}
 \caption{#2}\label{#3}
 \end{figure}}
 \newcommand{\inputplot}[3]{
 \special{ps: plotfile #1}
}
\newcounter{fig}   

\newcommand{\ze}{\kern 0.05em}

\newcommand{\beq}{\begin{equation}}
\newcommand{\eeq}{\end{equation}}
\newcommand{\beqs}{\begin{eqnarray}}
\newcommand{\eeqs}{\end{eqnarray}}

\newcommand{\be}{\begin{equation}}
\newcommand{\ee}{\end{equation}}
\newcommand{\bea}{\begin{eqnarray}}
\newcommand{\eea}{\end{eqnarray}}

\newcommand{\identity}{{\upright\rlap{1}\kern 2.0pt 1}}

\newcommand{\e}{\mbox{e}}

\def\theequation{\arabic{equation}}

\numberwithin{equation}{section}

\abstract{ 
A mechanism for circumventing the Mayo-Bekenstein no-hair theorem allows endowing four dimensional $(D=4)$ 
asymptotically  flat, spherical, electro-vacuum black holes with a minimally coupled $U(1)$-gauged  scalar field profile: $Q$-$hair$.
The scalar  field must be massive, self-interacting 
and obey a {\it resonance condition} at the threshold of (charged) superradiance.
We establish generality for this mechanism by endowing 
three
different types of 
static black objects with scalar hair, within a $D=5$  Einstein-Maxwell-gauged scalar field model: asymptotically flat
black holes
and black rings; and black strings
which asymptote to a Kaluza-Klein vacuum. These $D=5$ $Q$-hairy black objects share many of the features of their $D=4$ 
counterparts. 
In particular, the scalar field is subject to a resonance condition and possesses a 
$Q$-ball type potential.
For the  static black ring, 
 the charged scalar hair can balance it,
yielding solutions that are singularity free on and outside the horizon.
}

\preprint{ }

\title{$D=5$ static, charged black holes, strings and rings \\ with resonant, scalar $Q$-hair}

\author{{\large } 
{\large Y. Brihaye}$^{\dagger}$,
{\large C. Herdeiro}$^{\ddagger}$, 
and {\large E. Radu}$^{\ddagger}$
\\ \\
$^{\dagger}${\small 
Physique-Math\'ematique, Universite de
Mons-Hainaut, Mons, Belgium
	},
\\
$^{\ddagger}${\small  
Departamento de Matem\'atica  da Universidade de Aveiro and
 Center for Research and Development in Mathematics and Applications (CIDMA),
}
\\
{\small
 Campus de Santiago, 3810-183 Aveiro, Portugal}
 }

\begin{document}
 

\section{Introduction}
Influential results from the last three decades of the 20th century created a narrative that electrovacuum black holes (BHs) cannot support scalar ``hair"~\cite{Bekenstein:1996pn} if: 1) the scalar model is physical, $i.e.$ it obeys appropriate energy conditions, and 2) only couples minimally to both the gravitational and electromagnetic fields - see~\cite{Herdeiro:2015waa} for a review.\footnote{Dropping either assumption there are many models where BHs with scalar hair  occur, see $e.g.$~\cite{Herdeiro:2015waa,Sotiriou:2015pka}, for instance violating the weak or dominant energy condition  or allowing non-minimal couplings of the scalar field to the electromagnetic field or the geometry.} Two of the most influential theorems establishing the inexistence of scalar hair in these conditions, for uncharged and electrically charged BHs, respectively, were established by Bekenstein~\cite{Bekenstein:1972ny} and Mayo and Bekenstein~\cite{Mayo:1996mv}.

In the last decade, however, this narrative was debunked. Firstly,  it became clear that even for simple, physical, minimally coupled scalar models there is a generic mechanism allowing scalar hair around \textit{rotating} (either neutral or electrically charged) BHs. This mechanism relies on a \textit{synchronization condition}~\cite{Herdeiro:2014goa,Herdeiro:2014ima}. Physically it means no scalar energy flux exists through the horizon, allowing equilbrium between the scalar environment and the trapped region. Mathematically, the scalar field circumvents one innocuous looking hypothesis (symmetry inheritance~\cite{Smolic:2015txa}) of the aforementioned Bekenstein theorem. This mechanism has proved to be quite universal, applying (say) to both neutral and electrically charged rotating BHs, with different asymptotics, dimensions and horizon topologies - see $e.g.$ 
\cite{Dias:2011at,Brihaye:2014nba,Herdeiro:2015kha,Herdeiro:2015tia,Delgado:2016jxq,Herdeiro:2017oyt,Wang:2018xhw,Kunz:2019bhm,Delgado:2019prc}, as well as to other spin fields~\cite{Herdeiro:2016tmi,Santos:2020pmh}.

 In some cases, the BHs with synchronized hair bifurcate from the bald solutions. This occurs when the latter admit a linear version of the BH hair, called \textit{stationary scalar clouds}~\cite{Hod:2012px,Hod:2014baa,Benone:2014ssa,Hod:2016lgi}, that exist at the threshold of the superradiant instability of spinning BHs~\cite{Brito:2015oca}. This is the case for the paradigmatic Kerr solution of General Relativity; it means that creating synchronized hair is a natural dynamical process for Kerr BHs in the presence of such bosonic fields, as shown by  numerical evolutions \cite{East:2017ovw,Herdeiro:2017phl}  - see also the discussion in~\cite{Herdeiro:2022yle}.

The situation for \textit{charged} (non-spinning) BHs has both some some important similarities and some important differences as compared to that of spinning BHs. On the one hand, an analogous condition to the aforementioned synchronization condition -- which in essence establishes the possibility of an equilibrium between the hair and the horizon -- is possible for charged BHs; in this context it is called {\it  resonance condition}. Explicitly, it means that 
\begin{eqnarray}
\label{cond}
w=g_s V\big |_{{\cal H}}, 
\end{eqnarray}
where $w$ is the scalar field frequency, 
$g_s$ is the gauge coupling constant and 
$V\big |_{{\cal H}}$ 
is the value of the electric gauge potential at the horizon.
One the other hand, albeit a version of superradiance (charged superradiance~\cite{Brito:2015oca}) occurs for charged BHs, there are no scalar stationary clouds around the paradigmatic Reissner-Nordstr\"om (RN) BHs, at the threshold of charged  superradiance~\cite{Hod:2012wmy,Hod:2013nn}, except some marginally bound states at extremality~\cite{Degollado:2013eqa}, for a massive gauged scalar field model. This provides an impediment for hairy BHs bifurcating from the RN solution, akin to those birfurcation from the Kerr solution, thus vindicating  the Mayo-Bekenstein theorem.

In an interesting development, however, it was shown in~\cite{Hong:2019mcj,Herdeiro:2020xmb,Hong:2020miv}
that static hairy  BH solutions in 
Einstein-Maxwell-gauged scalar field  (EMgS) theory \textit{do} exist, but only if the scalar field possesses both a mass term \textit{plus} self-interactions; the latter invalidate one hypothesis of the Mayo-Bekenstein theorem. 
 Additionally,  the no go result
 in Ref. \cite{Hod:2013nn}
is also circumvented, since the scalar field does not become infinitesimally small
($i.e.$ the non-linearities are always relevant).
That is,  the solutions in~\cite{Hong:2019mcj,Herdeiro:2020xmb,Hong:2020miv}
 are not zero modes of the superradiant instability
and instead
can be viewd as 
{\it non-linear gauged Q-clouds}, in the spirit of (test field) $Q$-ball like solitons around BHs, first considered in~\cite{Herdeiro:2014pka}.

The main purpose of this work is to address the generality of the mechanism unveiled in~\cite{Hong:2019mcj,Herdeiro:2020xmb,Hong:2020miv} for endowing static, charged BHs with resonant, scalar $Q$-hair.  
To do so, we are gone consider the arena of higher dimensional BHs, which is more generous in terms of the possible black objects, even in vacuum and electrovacuum~\cite{Emparan:2008eg}. In four spacetime dimensions  $(D=4)$ for asymptotically flat BH solutions
only a spherical horizon topology is allowed. But
as the dimension increases, the phase structure
of the
 possible General Relativity 
(GR) 
black objects
becomes increasingly intricate and diverse 
\cite{Emparan:2008eg}.
 In this work we shall consider
three different types of static 
$D=5$
solutions,  corresponding to
 BHs, Black Rings (BRs) and Black Strings (BSs).
The BHs and BRs approach asymptotically a five dimensional
Minkowski spacetime background being
distinguished by their different horizon topology,
while the BSs are solutions in a Kaluza-Klein (KK) theory.
This allows us to attempt the construction of different types of BHs with resonant, gauged, $Q$-hair.

We shall focus on a complex massive scalar field with  ($Q$-ball type) quartic plus hexic self-interactions and
 establish that 
all qualitative results found 
in~\cite{Hong:2019mcj,Herdeiro:2020xmb,Hong:2020miv}
hold also for  the aforementioned
$D=5$ black objects.
Remarkably, in all cases studied, 
the emerging picture displays
common patterns.
 A crucial ingredient for 
the existence of such regular black objects
is  the  resonance condition  (\ref{cond}).
Moreover, this 
result
is independent on the self-gravity
effects:  
Maxwell-gauged scalar 
(MgS) solutions exist
already in the probe limit
($i.e.$ for a vacuuum black object background).
In all cases,
the coupling with gravity leads to a maximal horizon size
(which depends on the input parameters of the problem),
with a multi-branch structure of the space of solutions.
Moreover, owing to the pattern similarities for the $D=4$ and $D=5$ cases,
     we can reasonably expect that similar BHs with  gauged scalar hair 
     exist for any $D>5$.

This paper is organised as follows. 
In Section~\ref{section2} we present the EMgS models and propose a classification of the  solutions, based on their asymptotic behaviour.
The scaling symmetries and the generic features of the solutions 
 are also discussed. 
Section~\ref{section3} deals with solutions found by solving a set of
 ordinary differential equations (ODEs) -- the BHs and BSs.
In Section~\ref{section4}, we study the BRs and provide numerical evidence for
the existence of balanced  configurations, free of singularities on and outside the horizon.
We conclude in Section~\ref{section5} with a discussion and some further remarks.
The Appendix contains a brief review of three exact 
(static) solutions
in (pure) Einstein-Maxwell theory,
 corresponding to the $D=5$  
 RN BH, charged BR and charged BS.

\section{The general framework}
\label{section2}

\subsection{The action and field equations}
 
Working in $D=5$ spacetime dimensions, 
we consider the EMgS action: 
\begin{eqnarray}
\label{action}
 \mathcal{S}=\int d^5 x
\sqrt{-g}
\left(
\frac{R}{16\pi G}
-\frac{1}{4}F_{ab}F^{ab}
-D_a \Psi^*D^a \Psi-U(\left|\Psi\right|)
\right)~,
\end{eqnarray}
where $G$ is the gravitational constant, $R$ is the Ricci scalar associated with the
spacetime metric $g_{ab}$,
 $F_{ab} =\partial_a A_b - \partial_b A_a$ is the $U(1)$ field strength tensor,
and
\begin{eqnarray}
 D_a\Psi=\partial_a \Psi + ig_s A_a \Psi \ ,
\end{eqnarray}
is the gauge covariant derivative, with $g_s$ the gauge coupling constant.
  $ U(|\Psi|)>0$ denotes the potential of the complex scalar field $\Psi$,
whose mass
$\mu$
 is defined by  
\begin{equation}
\mu^2\equiv  \frac{\partial U}{\partial |\Psi|^2}\bigg|_{\Psi=0} \ .
\end{equation}

The EMgS field equations, 
obtained by varying the action with respect to the metric, scalar field and electromagnetic field, are, respectively,
\begin{eqnarray}
\label{E-eqs}
&&
R_{ab}-\frac{1}{2}g_{ ab}R=8 \pi G \left( T_{ab} ^{\rm (M)} +T_{ab}^{(\Psi)} \right), 
\\
\label{Ms-eqs} 
&&
D_{a}D^{a}\Psi=\frac{\partial U}{\partial\left|\Psi\right|^2} \Psi\ ,
\qquad 
\nabla_{a}F^{b a}=
i  g_s \big [\Psi^*(D^b \Psi)-(D^{b}\Psi^*) \Psi )   \big ]~
\equiv g_s j^b ~,
\end{eqnarray}  
with two different components in the total energy-momentum tensor 
\begin{eqnarray}
\label{tmunu} 
T_{ab}^{\rm (M)} 
&=&
F_a^{~c}F_{bc} - \frac{1}{4}g_{ab}F_{cd}F^{cd},~
\\
\nonumber
T_{ab}^{(\Psi)}
&=&
 D_a\Psi^* D_b\Psi 
+D_b\Psi^* D_a\Psi  
-g_{ab}  \left[ \frac{1}{2} g^{cd} 
 ( D_c \Psi^* D_d\Psi+
D_d\Psi^* D_c\Psi) +U(\left|\Psi\right|) \right]
 \ .
\end{eqnarray} 
This model is invariant under the local $U(1)$ gauge transformation 
\begin{eqnarray}
\label{gauge-transf}
\psi \to \psi e^{-i g_s \alpha}\ , \qquad A_\mu\to A_\mu +\partial_\mu \alpha \ ,
\end{eqnarray}
with $\alpha$ a real function of $x^a$.
Also, $j^a$ is the conserved current, $\nabla_a j^a=0$.

\subsection{Classes of solutions and global charges}

All geometries discussed in this work are static,
with a generic line element  
(see 
(\ref{metric-BH}),
(\ref{metric-BS}),
(\ref{metric-BR})):
\begin{eqnarray}
\label{metric-g}
ds^2=g_{tt}(\vec x)dt^2+g_{ij}(\vec x)dx^i dx^j \ ,
\end{eqnarray}
where $x^a=(t,\vec x)$ and $t$ is the time coordinate.
The matter fields
are of the form
\begin{eqnarray}
\label{matter}
 \Psi (\vec x,t)=\psi (\vec x) e^{-i wt} \ , \qquad  A(\vec x,t)=V(\vec x) dt \ ,
\end{eqnarray} 
with $\psi$
 and $V$
real functions and
$w$
the scalar field frequency.

There is also a  discrete symmetry
\begin{eqnarray}
\label{ds}
V\to -V, ~~g_s \to -g_s,
\end{eqnarray}
which allows us to consider the case $g_s \geq 0$ only.
 
Within this framework,
the only nonvanishing component of the conserved current is
\begin{eqnarray}
\label{jt}
j^t= {2(w- g_s V)\psi^2}g^{tt},
\end{eqnarray}
the associated Noether charge  (particle number) being
\begin{eqnarray}
\label{Noether}
Q_{N}= 
\int d^4x \sqrt{-g} j^t  , 
\end{eqnarray}
with the integral evaluated in the region outside the horizon.
In the generic case,
the  Noether charge provides a part of the total electric charge
 (as computed from the flux of the electric field
at infinity),
which,
from the second
equation in (\ref{Ms-eqs}),
can be written as the sum 
(with $Q_H$ the horizon charge)
 \begin{eqnarray}
\label{QeQ}
Q_{e}= Q_H+g_s Q_N \ , \qquad
{\rm where}~~Q_H=\oint_{{\cal H}}dS_i F^{ti}~.
\end{eqnarray}

\subsubsection{A Minkowski spacetime background}
Two classes of solutions discussed in this work,
the BHs and
the  BR  -- $cf.$ eqs.  (\ref{metric-BH}) and (\ref{metric-BR}) below, respectively --,
approach asymptotically a five dimensional
Minkowski spacetime background $\mathbb{M}^{1,4}$, with a line element
\begin{eqnarray}
\label{metric-Mink}
ds^2=-dt^2+ dr^2+r^2 d\Omega_3^2 \ , \qquad {\rm with} \qquad 
d\Omega_3^2=d\theta^2+\cos^2 \theta d\varphi_1^2+\sin^2\theta d\varphi_2^2 \ ,
\end{eqnarray}
where the range of $\theta$ is $0\leq\theta\leq \pi/2$ and  
  with $0\leq 
	(\varphi_1,\varphi_2)\leq 2 \pi$. 
	Also,  $r$ and $t$ correspond  to the radial and time
coordinates, respectively. 
The solutions possess a nonzero mass 
$M$
and 
an
electric charge
$Q_e$,
which are read off from the far field asymptotics of the metric function $g_{tt}$
and of the electric potential $V$, respectively,
\begin{eqnarray}
 g_{tt}=-1+\frac{8G M}{3\pi r^2}+\dots\ , \qquad 
V=\Phi-\frac{Q_e}{4 \pi^2 r^2}+\dots \ ,
\end{eqnarray}
with
$\Phi$
the chemical potential  
(for the
gauge discussed below).
The solutions possess an horizon which
can be of spherical topology, $S^3$ (BHs)
or of $S^2\times S^1$ topology (BRs).
The event horizon 
has a nonvanishing area
$A_H$.
The Hawking temperature
$T_H$
of all of the 
considered
 solutions is also nonzero, and can be computed from their surface
gravity.

In order to compare the pattern of
the hairy  solutions with
that of the known electrovacuum BHs,  
it is useful to consider
 reduced quantities,
with horizon area,
Hawking temperature and
electric charge normalized $w.r.t.$
the mass of the solutions
\begin{eqnarray}
\label{scale-BHR} 
a_H=\frac{3}{32}\sqrt{\frac{3}{ 2 \pi}}\frac{A_H}{(G M)^{3/2}}\ , \qquad 
t_H=4\sqrt{\frac{2\pi}{3}} T_H \sqrt{G M}\ , \qquad  
q=\frac{\sqrt{3G}}{4 \sqrt{\pi} }\frac{Q_e}{G M}~,
\end{eqnarray}
 with the coefficients 
chosen such that
$a_H=t_H=1$
in the Schwarzschild-Tangerlini limit, 
while $q=1$
for an extremal ($D=5$)  RN background -- 
see Appendix A.

For any horizon topology,
the solutions satisfy
the 1st law of thermodynamics\footnote{In which case the BRs are necessarily
balanced
\cite{Herdeiro:2009vd,Herdeiro:2010aq,Astefanesei:2009mc}.} 
\begin{eqnarray}
\label{1st}
dM= \frac{T_H}{4 G}dA_H +  \Phi dQ_e \ ,
\end{eqnarray}
and 
the Smarr relation:
\begin{eqnarray}
\label{Smarr} 
M=\frac{3}{2}T_H \frac{A_H}{4G}+  \Phi Q_e+ M_{(\psi)} ,
\end{eqnarray}
with $M_{(\psi)}$ the mass outside the horizon stored in the scalar field,
which, for the chosen gauge
(see the discussion in
Section~\ref{section2g})
takes the simple form 
\begin{eqnarray}
\label{Mint}
  M_{(\psi)}=  \int d^4 x\sqrt{-g} \left (\frac{1}{2} g_s V j^t -U(\psi) \right) \ ,
\end{eqnarray}
with the integral  
evaluated
in the region
 outside the horizon.

\subsubsection{A Kaluza-Klein spacetime background}

The second case 
corresponds to  black objects
approaching asymptotically four dimensional Minkowski spacetime times a circle,
$\mathbb{M}^{1,3}\times S^1$.
 We denote the compact direction as $z$, with an arbitrary periodicity $L$,
such that the background spacetime metric is 
\begin{eqnarray}
\label{metric-KK}
ds^2= -dt^2+ dr^2+r^2 d\Omega_2^2 + dz^2, 
\end{eqnarray}
with $d\Omega_2^2 $ the metric on a two-sphere. 

For any static spacetime which is asymptotically 
$\mathbb{M}^{1,3}\times S^1$
one can define a mass $M$, a tension ${\cal T}$,
and an electric charge $Q_e$, 
these quantities being encoded in the asymptotics of the metric potentials
\cite{Traschen:2001pb,Harmark:2003dg}
and of the electrostatic potential,
with
\begin{eqnarray} 
g_{tt}=-1+\frac{c_t}{r}+\dots \ , \qquad 
g_{zz}= 1+\frac{c_z}{r}+\dots \ , \qquad 
V=\Phi-\frac{1}{4\pi L}\frac{Q_e}{r}+\dots \ , 
\end{eqnarray}
and
\begin{eqnarray} 
M=\frac{L}{4G} (2c_t-c_z) \ , \qquad {\cal T}=\frac{1}{4G} (c_t-2c_z) \ .
\end{eqnarray}
One can also define a relative tension
$
n={{\cal T} L}/{M},
$
with
$n=1/2$ for a Schwarzschild BS, 
while
$0\leq n\leq 1/2$ for a BS in Einstein-Maxwell (EM) theory.
The BSs possess also an horizon area $A_H$
and a Hawking temperature $T_H$. 
As with the solutions with 
$\mathbb{M}^{1,4}$
asymptotics,
we define a set of 
reduced
quantities normalized $w.r.t.$ mass,
\begin{eqnarray} 
\label{scale-BS}
a_H=\frac{1}{16 \pi}\frac{A_H L}{(G M)^{2}}, \qquad 
t_H=8\pi T_H \frac{ G M}{L},  \qquad 
q=\frac{3\sqrt{G}}{4\sqrt{3\pi}}\frac{Q_e L}{G M}~,
\end{eqnarray} 
such that
$a_H=t_H=1$
is the Schwarzschild-BS limit, 
while $q=1$
for an extremal  BS solution in EM theory 
(see the discussion in Appendix A).

For Kaluza-Klein asymptotics, the 1st law of thermodynamics contains
an extra-term, 
 \begin{eqnarray}
\label{1st-KK}
dM= \frac{T_H}{4 G} dA_H +  \Phi dQ_e +{\cal T} dL,
\end{eqnarray}
and so does the Smarr relation,
\begin{eqnarray}
\label{Smarr-BS}
M=\frac{3}{2} \frac{T_H}{4 G}A_H 
+\Phi Q_e
+ \frac{1}{2} {\cal T} L
+ M_{(\psi)} ,
\end{eqnarray}
 with $M_{(\psi)} $
still given by (\ref{Mint}).

\subsection{The potential and scaling properties }
 
For a quantitative study of the solutions, we need to specify the expression
of the potential $U(\left|\Psi\right|)$. 
Following the previous $D=4$ work, 
the results in this paper are for a 
potential which is the sum of a mass term plus quartic and sextic self-interactions:
\begin{eqnarray}
\label{potential}
U(|\Psi|)=\mu^2 |\Psi|^2-\lambda |\Psi|^4+\nu |\Psi|^6~,
\end{eqnarray}
where 
$\mu$ is the scalar field mass and 
$\lambda,\nu$ are positive parameters
(with $\lambda^2<4\mu^2 \nu$
for a strictly positive potential,
$U(|\Psi|)>0)$.
As with the $D=4$ case,
 the presence of higher order terms in the scalar potential
appears to be mandatory, and we have failed
to find hairy solutions
for a scalar field with a mass term only.
  
In the numerics,   
it is useful to work with a set of scaled, dimensionless:  
$i)$ model input parameters; 
$ii)$ matter functions;  
and
$iii)$ a dimensionless  radial coordinate.
These are
defined as 
(denoted with overbar in what follows):
\begin{eqnarray}
\label{scale}
w=\bar w \mu, \qquad 
g_s=\bar g_s \sqrt{\lambda}, \qquad 
V=\bar V \frac{\mu}{\sqrt{\lambda}}, \qquad 
\Psi=\bar \Psi \frac{\mu}{\sqrt{\lambda}}
 \qquad {\rm and} \qquad 
r=\frac{\bar r}{\mu}.
 \end{eqnarray}  
This scaling reveals the existence of three input dimensionless parameters
 \begin{eqnarray}
\alpha^2=\frac{4\pi G \mu^2}{\lambda}, \qquad 
\beta^2=\frac{\nu \mu^2}{\lambda}, \qquad 
  \bar g_s=g_s \sqrt{\lambda} \e  e~,
 \end{eqnarray} 
 which characterize a given model. 

Under the transformation (\ref{scale}),
several quantities of interest behave as
\begin{eqnarray}
M=\frac{\bar M}{\mu^2 G}, \qquad 
Q_e=\frac{\bar Q_e}{\mu \sqrt{\lambda}}, \qquad 
\Phi=\bar \Phi \frac{\mu}{\sqrt{\lambda}} \qquad 
A_H=\frac{\bar A_H}{\mu^3}, \qquad 
T_H=\bar T_H \mu \ .
\end{eqnarray} 

The numerics is done with the scaled quantities and functions, 
and dimensionless parameters.
With these conventions, the Einstein equations 
solved numerically are
$R_{\mu\nu}-\frac{1}{2}g_{\mu\nu} {  R}= 2\alpha^2~T_{\mu\nu}$,
while 
 the scaled scalar potential --  for the ansatz (\ref{matter}) -- 
 is 
$
U(\bar \psi)=\bar \psi^2-\bar \psi^4 +\beta^2 \bar \psi^6.
$
However, to simplify the picture, 
  we shall ignore the overbar
	in the plots for 
$V$ and $\psi$.
Also, for the sake of clarity all equations displayed 
in what follows are given
in terms of dimensionful variables.

\subsection{The bound state condition
and
fixing the gauge}
\label{section2g}
For large-$r$, the deviation from the background geometry
can be neglected to leading order in the scalar field equation.
This leads to the following
 asymptotic expression of the scalar field 
\begin{eqnarray}
\label{asympt-scalar}
\psi(r) \sim  \frac{e^{-\mu_{\rm eff} r}}{r^{3/2}}+\dots,~~
{\rm for~BHs~and~BRs},~~{\rm and}
~~
\psi(r) \sim \frac{e^{-\mu_{\rm eff} r}}{r }+\dots,~~
{\rm for~BSs},
\end{eqnarray}
where we define
\begin{eqnarray}
 \mu_{\rm eff} =\sqrt{\mu^2-(w-g_s \Phi)^2}.
\end{eqnarray}

A simple inspection of the equations reveals that
the scalar field frequency
$w$
enters always in the combination
$w-g_s V$. 
Thus the model still possesses the residual gauge symmetry
\begin{eqnarray}
\label{rgs22}
w\to w+\gamma, \qquad V\to V+s\gamma/g_s~,
\end{eqnarray}
(with $\gamma$
a real number)
which should be fixed in numerics.
Following~\cite{Herdeiro:2020xmb},
we work in a gauge with a vanishing electric potential 
at the horizon, 
 \begin{eqnarray}  
\label{gauge}
 V\big |_{{\cal H}}=0 .
\end{eqnarray} 
Then the regularity condition
 (\ref{cond}) 
implies $w=0$, $i.e.$ a real scalar field.
As such, the matter Lagrangian of the model 
can be written in the suggestive form
\begin{eqnarray}
\label{Ln}
\mathcal{L}_m=
-\frac{1}{4}F_{ab}F^{ab}
-\partial_a \psi \partial^a \psi-g_s^2 A_a A^a \psi^2-U(\psi),
\end{eqnarray} 
with the vector potential
acquiring a local mass.  Note, however, that 
 $\psi$
is not a Higgs field,
since it vanishes asymptotically.
  
It is also interesting to investigate the status of
the  Mayo-Bekenstein no-go result
\cite{Mayo:1996mv} 
for the considered $D=5$ configurations.
The starting point is  
the equation for the electric potential $V$,
which, in the chosen gauge is
\begin{eqnarray}
\nabla_a F^{ta}=-2g^{tt}g_s \psi^2V ~. 
\end{eqnarray}
Multiplying it by $V$ and integrating by parts yields
\begin{eqnarray}
\label{neq}
\oint_{\infty} dS_i  V F^{ti} =\Phi Q_e = -\int d^4 x \sqrt{-g} 
g^{tt}
 \left[
g^{ab} \partial_a V \partial_b V+2g_s V^2 \psi^2
\right],  
\end{eqnarray}
where we have used condition (\ref{gauge}).
Since the $r.h.s.$
of the above expression is strictly positive,
this implies that the solutions necessarily have nonzero $Q_e$
{\it and} $\Phi$.
If, as assumed by Mayo and Bekenstein, there is no mass term
in the potential
$U(\psi)$
(or $\mu<g_s \Phi$) 
the scalar field would 
possess wave-like 
asymptotics   
(see the eq. (\ref{asympt-scalar})),
and thus one is forced to impose $\Phi=0$.
Then (\ref{neq}) implies that the scalar field necessarily vanishes.

\subsection{Generic features of the solutions}

In Sections 
\ref{section3} and 
\ref{section4},
we shall present the equations and the boundary conditions,  
together with a study of the solutions for the different hairy black objects we shall construct.
Without entering into  details,
here we summarize  their common basic features.

\begin{itemize}

\item
Within the considered framework, the numerical
problem contains
five
input parameters
\begin{eqnarray}
\{
\alpha,\beta,e; \Phi; r_H
\}; 
\end{eqnarray}
the first three correspond
to constants of the model,
while $\Phi$
is the  asymptotic value of the electric potential
 and $r_H$
is the horizon radius.
For the BRs, there is one more input parameter,
 the ring's radius $R$.
The construction reduces to solving a set of
ODEs (for BHs and BSs) or Partial Differential Equations (for BRs).
The quantities of 
interest 
are computed from the numerical output.

\item
 The limit of a vanishing scalar field 
$\psi=0$
is a consistent solution of the model.
In this case we recover 
the known
(static) BHs, BRs and BSs in EM theory,
whose basic properties are reviewed in Appendix A.

\item
Apart from the EM solutions, 
there are black objects with 
gauged scalar hair. 
However,
the scalar field
 does not emerge as a zero mode 
of an electrovacuum solution.
That is, the scalar field never trivializes,
with the necessary existence of 
non-linear terms in the scalar potential.

\item
The solutions  
satisfy the 
{\it resonance condition}
(\ref{cond}),
which emerges
from assuming the existence of
a power series expansion of the solutions 
close to the horizon
together with regularity conditions.
 Also, working in a gauge with $w=0$,
the solutions satisfy the {\it bound state} condition
\begin{eqnarray}
\label{bound}
g_s \Phi \leq \mu.
\end{eqnarray}

\item
Non-linear gauged $Q$-clouds exist 
already
in the decoupling limit  of the model, 
$i.e.$
when ignoring the backreaction of the scalar  and
Maxwell field on the geometry
and
 solving MgS field equations
on a fixed background
which corresponds to
the
Schwarzschild-Tangherlini BH, 
a vacuum BR and
a Schwarzschild BS, respectively.

 \item
The solutions of the full EMgS model 
display a complicated pattern,
with a maximal horizon size 
and different branches of solutions.
The branch of fundamental solutions
possesses a well defined 
horizonless
limit corresponding to charged $Q$-balls 
(for a $\mathbb{M}^{1,4}$ background)
and $Q$-vortices 
(in the $\mathbb{M}^{1,3}\times S^1$ case).
 
\item
All reported solutions have a non-zero Hawking temperature,
while the model is unlikely to possess
extremal BH solutions\footnote{
One hint in this direction is the absence,
at least for an $S^3$ horizon topology,
 of the usual  attractor solutions, 
$i.e.$ generalizations of the Bertotti-Robinson solution,
with a metric $AdS_3\times S^2$. }.

\end{itemize}

\section{ Co-dimension one solutions.
  Black Holes and Black Strings}
\label{section3}

The BHs and BSs discussed in this work solve a set of ODEs
with suitable boundary conditions.
Since the treatment of the numerical problem
together with the unveiled picture is rather similar,
we report them together in what follows.

\subsection{The Ansatz and equations}

\subsubsection{Black holes }

In the numerical study of the
spherically symmetric
 solutions, it is convenient to use the following metric Ansatz:
\begin{eqnarray}
\label{metric-BH}
ds^2=-N(r)\sigma^2(r)dt^2+\frac{dr^2}{N(r)}+r^2 d\Omega_3^2 \ , 
\end{eqnarray}
while the matter functions $\psi$ and $V$
depend  on the radial coordinate $r$ only. 
Then the corresponding field equations\footnote{
In all three cases there is also an extra-constraint
equation (two eqs. for BRs), 
which is not solved directly, being used to
check the consistency of the numerical results. },
as resulting from (\ref{E-eqs}),  (\ref{Ms-eqs})
 read 
\begin{eqnarray}
&&
\nonumber
N'+\frac{2}{r}(N-1)+ \frac{16\pi G}{3} r
\left[
\frac{V'^2}{2\sigma^2}+N\psi'^2+U(\psi)+\frac{(w-g_s V)^2}{N\sigma^2}\psi^2
\right]=0,
\\
\label{eqm}
&&
\sigma'=\frac{16\pi G}{3}  r \sigma
\left[
\psi'^2+\frac{(w-g_s V)^2\psi^2}{N^2 \sigma^2}
\right],
\\
&&
\nonumber
V''+\left(\frac{3}{r}-\frac{\sigma'}{\sigma}\right)V'+\frac{2q_s (w-g_s V)\psi^2}{N}=0,
\\
&&
\nonumber
\label{eqpsi}
\psi''+
\left(
\frac{3}{r}+\frac{N'}{N}+\frac{\sigma'}{\sigma}
\right)\psi'
+\frac{ (w-g_s V)^2 \psi}{N^2\sigma^2}
-\frac{1}{2N}\frac{dU}{d\psi}=0~.
\end{eqnarray}
The non-vanishing
components of the energy-momentum tensor are
\begin{eqnarray}
\nonumber 
&&
T_r^{r(M)}=-\frac{V'^2}{2\sigma^2}, \qquad
T_r^{r(\Psi)}=N\psi'^2+\frac{(w-g_s V)^2 \psi^2}{N\sigma^2}-U(\psi) \ ,
\\
\label{Tij}
&&
T_\Omega^{\Omega(M)}= 
\frac{V'^2}{2\sigma^2} \ ,  \qquad
T_\Omega^{\Omega(\Psi)} =
-N\psi'^2+\frac{(w-g_s V)^2\psi^2}{N\sigma^2}-U(\psi),
\\
\nonumber
&&
T_t^{t(M)}=-\frac{V'^2}{2\sigma^2} \ ,  \qquad
T_t^{t(\Psi)}=-N\psi'^2-\frac{(w-g_s V)^2\psi^2}{N\sigma^2}-U(\psi)~,
\end{eqnarray}
with $\Omega=(\theta,\varphi_1,\varphi_2)$.
The horizon is located at $r=r_H>0$
 with
$N(r)\sim (r-r_H)$ as $r\to r_H$,
while the metric function $\sigma(r)$
remains nonzero (and finite) in the same limit.
The Hawking temperature 
and the event horizon area of the solutions are found from the horizon data,
\begin{eqnarray}
T_H=\frac{1}{4\pi}N'(r_H)\sigma(r_H) \ ,  \qquad A_H=2\pi^2 r_H^3~.
\end{eqnarray}

We assume the existence of a  power series of
the solutions in $(r-r_H)$ close to the horizon. 
Then the  finiteness of the energy-momentum
tensor (\ref{Tij})
(or of the current density (\ref{jt}))
implies the following condition
\begin{eqnarray}
\psi(r_H) (w - g_s V(r_H)) =0 \ ,
\end{eqnarray}
which can be satisfied by taking (\ref{cond})
or by taking  $\psi(r_H)=0$.
However, the latter  choice 
implies that  
the derivatives of the scalar field 
vanish order by order in the power series expansion
close to the horizon; 
 that is, the scalar field trivializes.
 Thus the only reasonable solution is the resonance condition 
(\ref{cond}).

\subsubsection{Black strings }
The line element in this case contains three unknown functions, with\footnote{After a Kaluza-Klein reduction,
the BSs possess a $D=4$ description as 
(spherically symmetric)
BHs in a EMgS model with an extra dilaton
field, whose value is given by the metric component
$g_{zz}=a(r)$.}
\begin{eqnarray}
\label{metric-BS}
ds^2=-b(r) dt^2 + \frac{dr^2}{f(r)}+r^2 d\Omega_2^2+a(r)dz^2,
\end{eqnarray}
with $\psi$ and $V$
functions of $r$
only.
 Then
the EMgS equations read
\begin{eqnarray}
&&
\nonumber
 f'+\frac{2(f-1)}{r}
+\bigg(\frac{a'}{a}+\frac{b'}{b}\bigg)f
+\frac{32 \pi G}{3}r
\left[
\frac{f}{2b}V'^2
+U(\psi)
\right]=0 \ ,
\\
&&
\nonumber
a'\bigg(b+\frac{1}{4}rb'\bigg)
+\bigg[b'+\frac{b}{r}\bigg(1-\frac{1}{f}\bigg)\bigg]a
+8\pi G r a
\left[
\frac{1}{2}V'^2 -b\psi'^2
+\frac{b}{f}U(\psi)
-\frac{(w-g_s V)^2 }{f}\psi^2
\right]=0 \ ,
\\
\nonumber
&&
b''
+ \left[
\frac{1}{r}\bigg(1+\frac{1}{f}\bigg)
-\frac{b'}{b}
 \right] b'
+\frac{16 \pi G}{3}
\left[
-\frac{6(w-g_s V)^2}{f}\psi^2
-2(1+\frac{rb'}{4b})V'^2
+\frac{2b-r b'}{f}U(\psi)
\right]
=0 \ ,
\\ 
&&
\label{eqf}
\psi''+\frac{1}{2}
\left(
\frac{4}{r}+\frac{a'}{a}+\frac{b'}{b}+\frac{f'}{f}
\right) \psi'
+\frac{(w-g_s V)^2}{b f}\psi
-\frac{1}{2}\frac{dU}{d\psi}\frac{\psi}{f}=0 \ ,
\\
&&
\nonumber
V''+\frac{1}{2}
\left(
\frac{4}{r}+\frac{a'}{a}+\frac{b'}{b}+\frac{f'}{f}
\right) V'
+\frac{2g_s(w-g_s V)\psi^2}{f}=0~.
\end{eqnarray}
The non-vanishing
components of the energy-momentum tensor
are
\begin{eqnarray}
\nonumber
&&
T_r^{r(M)}=-\frac{fV'^2}{2b} \ , \qquad 
T_r^{r(\Psi)}=f\psi'^2+\frac{(w-g_s V)^2\psi^2}{b}-U(\psi)\ ,~~
\\
\nonumber
&&
T_\Omega^{\Omega(M)}= \frac{fV'^2}{2b}\ , \qquad 
T_\Omega^{\Omega(\Psi)} =-f\psi'^2+\frac{(w-g_s V)^2\psi^2}{b}-U(\psi)\ ,~~
\\
\label{Tij-s}
&&
T_t^{t(M)}=-\frac{fV'^2}{2b}\ , \qquad 
T_t^{t(\Psi)}=-f\psi'^2-\frac{(w-g_s V)^2\psi^2}{b}-U(\psi)~,
\\
\nonumber
&&
T_z^{z(M)}=\frac{fV'^2}{2b} \ , \qquad 
T_z^{z(\Psi)}=-f\psi'^2+\frac{(w-g_s V)^2\psi^2}{b}-U(\psi)~.
\end{eqnarray}

\begin{figure}[ht!]
\begin{center}
\includegraphics[height=.34\textwidth, angle =0 ]{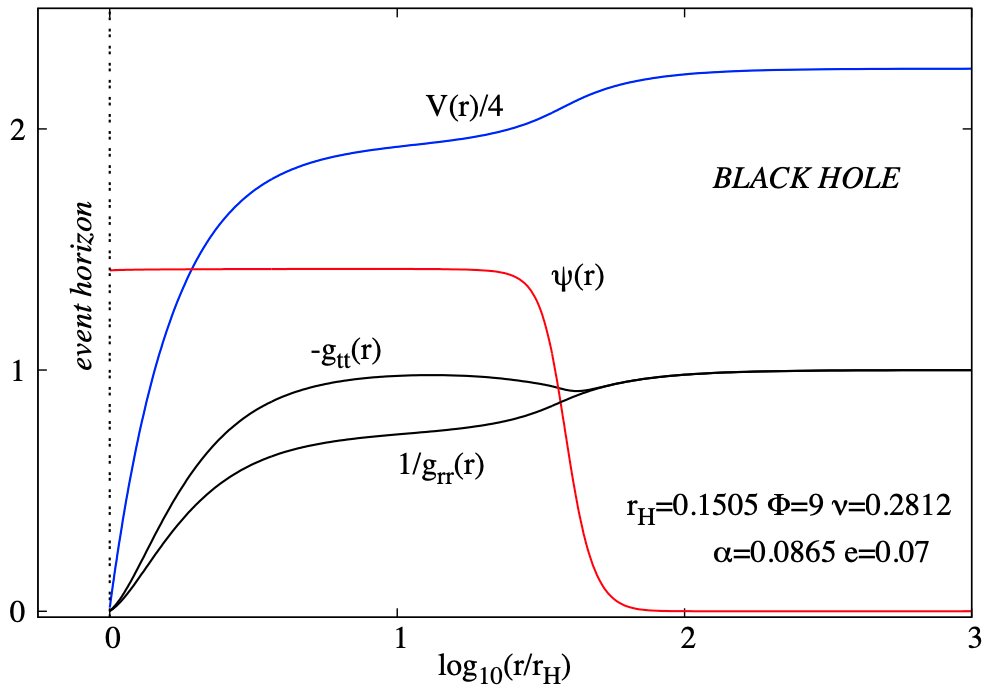}
\includegraphics[height=.34\textwidth, angle =0 ]{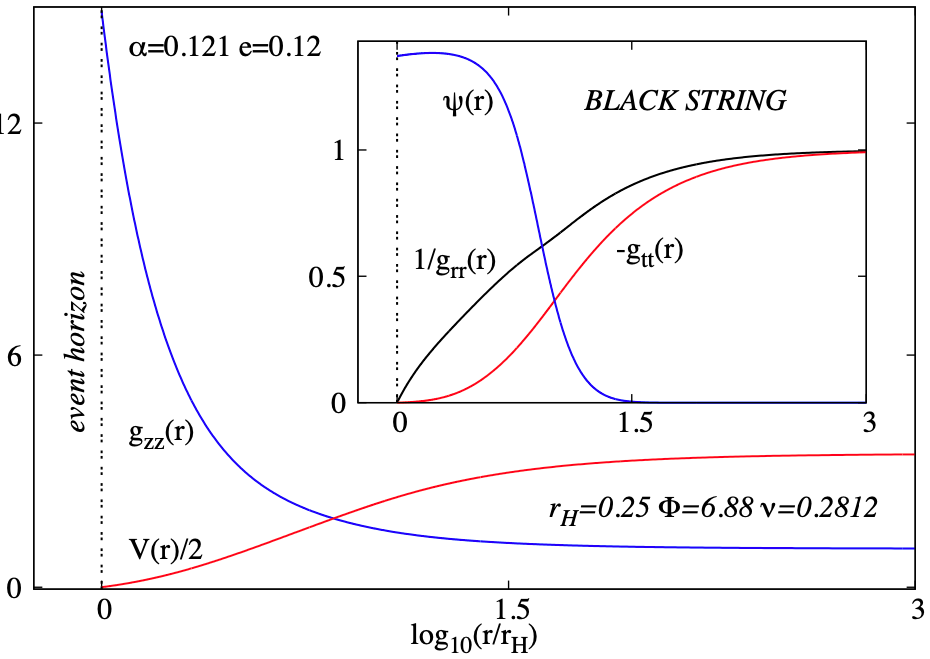}
\end{center}
\caption{
Profile functions of a typical BH (left)
and BS (right) 
are shown
as functions of the radial coordinate. 
}
\label{profile}
\end{figure}
The horizon is again located for some $r=r_H>0$,
where $f(r_H)=b(r_H)=0$,
while $a(r_H)>0$.
The horizon area and Hawking temperature are
\begin{eqnarray}
A_H=4\pi r_H^2 L \sqrt{a(r_H)}\ , \qquad  T_H=\frac{1}{4\pi }\sqrt{f'(r_H) b'(r_H)}~.
\end{eqnarray}

As with the BHs, the resonance condition (\ref{cond})
necessarily emerges
when assuming the existence of a power series
expansion of the solutions close to the horizon.

\subsection{The results}
In the absence of closed form solutions,
the set of four (five) ODEs
(\ref{eqm})
 (and (\ref{eqf}), respectively)
are solved numerically,
by using a
professional  solver
that employs
a collocation method for boundary value
ODEs equipped with an adaptive mesh selection procedure 
\cite{colsys}. 
Typical mesh sizes include few hundred points,
the relative accuracy of the solutions being
around $10^{-10}$.
The boundary conditions we have imposed are:
\begin{eqnarray}
\nonumber
&&
{\bf BHs}:~
N \big |_{r=r_H}=
V \big |_{r=r_H}=
\psi \big |_{r=\infty}=0,~
\bigg[ N' \psi'-\frac{1}{2}\frac{dU}{d\psi} 
\bigg]_{r=r_H}=0,~
N\big |_{r=\infty}=1,~
V \big |_{r=\infty}=\Phi,~
\\
&&
{\bf BSs}:~
f|_{r=r_H}=b|_{r=r_H}=V|_{r=r_H}=0,~~
\bigg[ \bigg(\frac{a'}{a}+\frac{b'}{b}\bigg) \psi'-\frac{1}{2}\frac{dU}{d\psi} 
\bigg]_{r=r_H}=0,~~
\\
&&
\nonumber
{~~~~~~~~~}
f\big |_{r=\infty}=b\big |_{r=\infty}=1,~~
V \big |_{r=\infty}=\Phi,~~
\psi \big |_{r=\infty}=0,~~ a \big |_{r=\infty}=1,~~ 
\end{eqnarray}
which result from
a study of 
the near horizon and far field approximate form of the solutions.

\medskip
The profiles of a typical BH/BS solution 
is shown in Figure \ref{profile}.
One can see that in both cases the matter functions 
monotonically interpolate between the horizon and infinity.
Solutions with nodes for both 
$V$ and $\psi$ do also exist;
in particular,
we have found numerical evidence for 
the existence of
$D=5$ 
generalizations of the four-dimensional
'wavy' hairy  BH discussed in~\cite{Brihaye:2021phs,Brihaye:2021mqk}.
However, in this work we shall restrict 
ourselves to the study
of nodeless configurations.


\begin{figure}[ht!]
\begin{center} 
\includegraphics[height=.34\textwidth, angle =0 ]{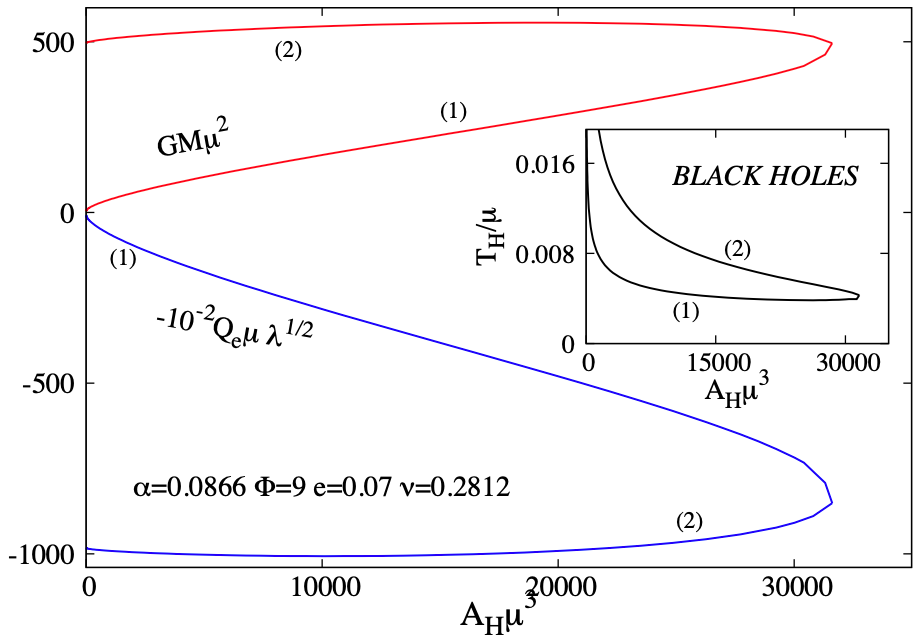}
\includegraphics[height=.34\textwidth, angle =0 ]{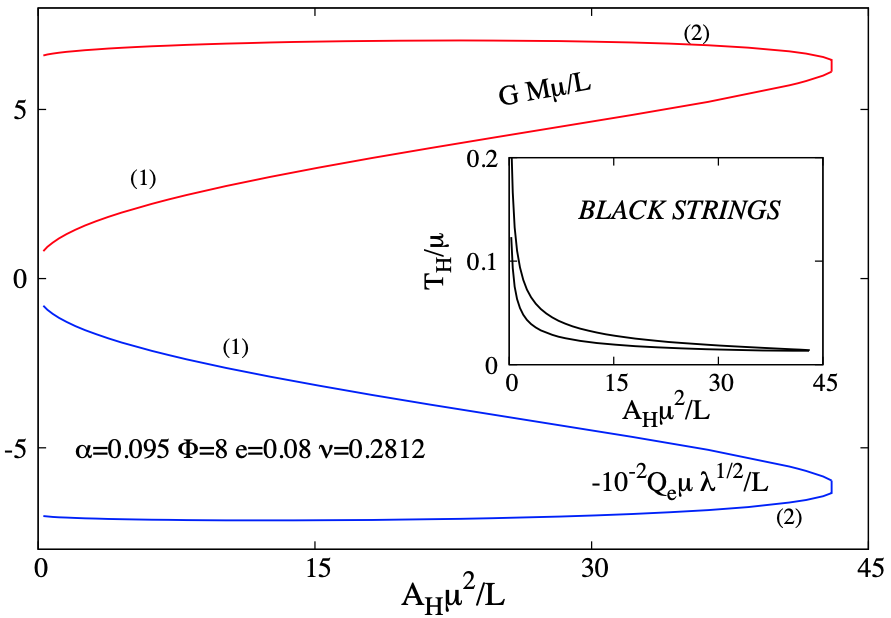}
\end{center}
\caption{ 
The mass, electric charge and Hawking temperature of solutions
are shown as a function of event horizon area
for BHs and BSs.
The quantities are shown in the natural units of the model.
}
\label{AHMTH}
\end{figure}

The complete classification of the solutions 
in the space of physical parameters
$(\alpha,\beta,e)$
 is a considerable
task which is not the goal of this work.
Here, for both 
BHs and BSs (and also for BRs in the next Section)
 we shall  present illustrative results 
for
a selected set 
$(\alpha,\beta,e)$
in each case, as a proof of concept\footnote{
We emphasize, however, that
the equations  
 have been solved 
for other choices of theory parameters
$(\alpha,\beta,e)$ and few more values of $\Phi$.}.
Also, the BH/BS horizon size is  
varied for a fixed value of the 
electrostatic potential  $\Phi$ --
that is, we study solutions in a grand canonical ensemble --, 
and we do not consider the case
 with a fixed electric charge.

When reporting on the properties of the solutions,
it is natural to start with their horizonless limit, approached as $r_H\to 0$.
This nontrivial 
limit 
exists due to the scalar field interaction, 
being absent (or pathological) for the pure Einstein-Maxwell case.
It  corresponds to 
gauged $Q$-balls   (for a $\mathbb{M}^{1,4}$ background)
and gauged $Q$-vortices 
(for a $\mathbb{M}^{1,3}\times S^1$ background).
These are $D=5$ natural counterparts of the
four dimensional
EMgS solitonic
 solutions 
reported in  
\cite{Jetzer:1989av,Jetzer:1993nk,Pugliese:2013gsa,Prikas:2002ij,Brihaye:2014gua},
and appear to share with them all basic properties.
In particular, 
when fixing the input parameters of the model $(\alpha,\beta,e)$,
the horizonless solutions exist again for a finite range of the 
parameter $\Phi$,
the upper limit being
fixed by the bound state condition (\ref{bound}),
while $\Phi_{\rm min}$ is nonzero and results from numerics.

\begin{figure}[ht!]
\begin{center} 
\includegraphics[height=.34\textwidth, angle =0 ]{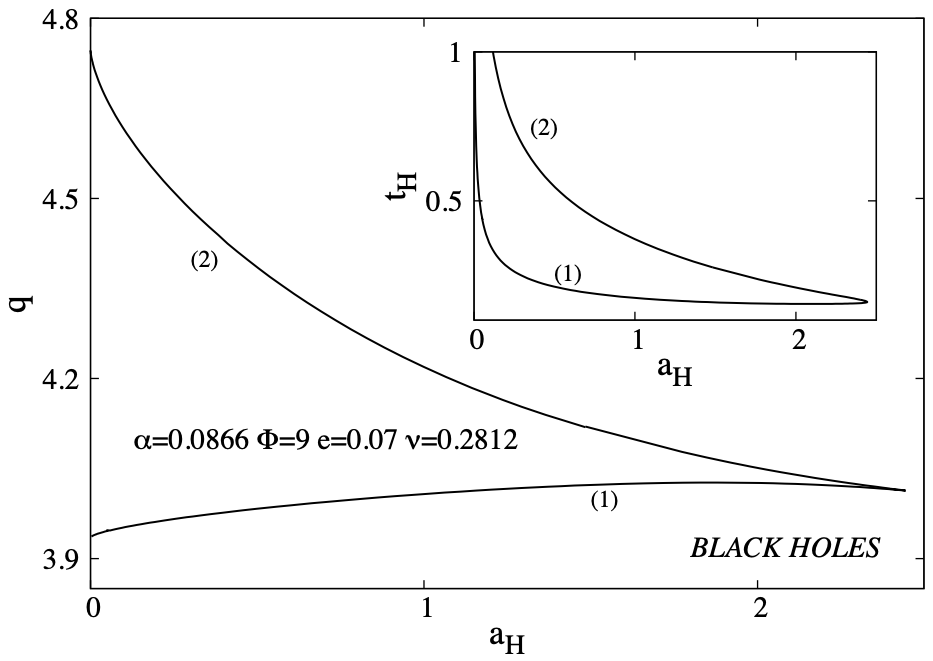}
\includegraphics[height=.34\textwidth, angle =0 ]{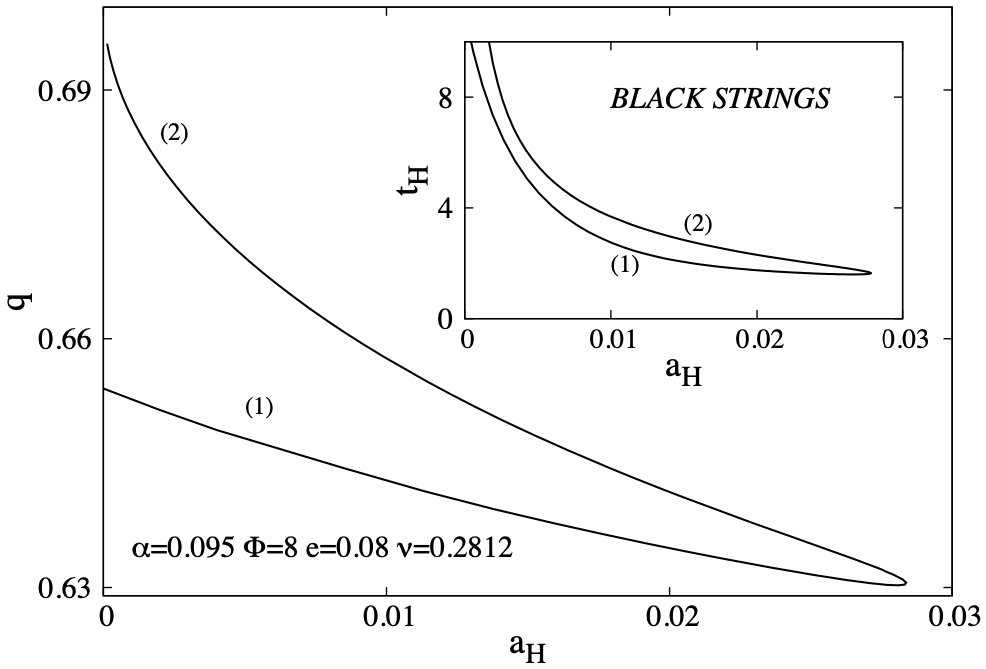}
\end{center}
\caption{
The reduced electric charge  and Hawking temperature
are shown as a function of reduced horizon area for BHs and BSs. 
The quantities are shown in units set by the mass of the solutions.
}
\label{aHtH}
\end{figure}

Any gauged $Q$-ball/$Q$-vortex solution appears to possess BH generalizations.
Given the parameter 
$(\alpha,\beta, e;\Phi)$,
the BHs and BSs are found
 by  slowly increasing from zero the value of  event horizon radius.
As shown in 
Figures \ref{AHMTH},  
\ref{aHtH}, 
the solutions with a fixed chemical potential $\Phi$  
exist up to a maximal BH size, as specified by the event horizon area $A_H$.

Along this fundamental branch 
(denoted with (1) in the plots), 
both the mass and the electric charge increase\footnote{In the BS case,
a similar behaviour is found for the string tension
${\cal T}$
(not displayed here).
However, it is interesting to note that
the relative tension $n$
does not vary significantly for the displayed set;
 $e.g.$ the vortex-limiting solutions have
 $n\simeq 0.754$ for the fundamental branch and 
 $n\simeq 0.773$ on the second branch. 
} with $A_H$.
At the same time, the 
Hawking temperature and the
value of the scalar field at the horizon decrease.
As $A_H\to A_H^{\rm (max)}$, a secondary branch emerges, with a backbending in $A_H$.
However, the end state of this secondary branch
(denoted with (2)
in Figures
\ref{AHMTH},  
\ref{aHtH}),
 depends on the value of the electrostatic potential $\Phi$.
The behaviour reported in~\cite{Herdeiro:2020xmb}
for $D=4$ BHs 
is recovered for large enough values of 
 the electrostatic potential,
and this secondary branch stops to exist for a  
nonzero $A_H$, where
the numerics becomes increasingly challenging, the scalar
field being confined in a region close to the horizon.
In Figures
\ref{AHMTH},  
\ref{aHtH}
 we present results 
for a different limiting behaviour
 which is found for a different range of 
$\Phi$,
with a gauged 
$Q$-ball/$Q$-vortex
 being approached as $A_H \to 0$
also for  the secondary branch of solutions\footnote{
This behaviour occurs for some range of $\Phi$
only.}.

The pattern found when showing the horizon area,
electric charge and temperature scaled $w.r.t.$
the total mass is somehow similar,
see Figure \ref{aHtH} 
(note, however, the
different behaviour on the fundamental branch for BHs and BSs).
Also, when comparing the results in Figures
\ref{aHtH}
and  
\ref{RNBHBS} (the latter is for the electrovacuum BHs and BSs),
one notices that the hairy solutions 
exhibit a very different pattern as compared to the
corresponding electrovacuum solutions.

Finally, we have found that
for given $(\alpha,\beta)$ and a fixed value of  $ \Phi$,
the solutions exist for a finite range of the 
(scaled) gauge coupling constant
$e$, only.
Moreover,
as in the solitonic limit,
the BHs/BSs exist for a finite range of coupling constant	$\alpha$, only.
The limit $\alpha \to 0$
is of special interest, 
and corresponds
to solving
the MgS field equations
on a fixed  background, which
corresponds to the
Schwarzschild-Tangherlini BH
($N=1-(r_H/r)^2$,
$\sigma=1$
in (\ref{metric-BH})), 
or  a Schwarzschild BS
($f=b=1-r_H/r$,
$a=1$
in (\ref{metric-BS})).
This 'probe limit' is technically simpler, while
the solutions capture already some of the basic feature
of the backreacting configurations.
For example, these MgS solutions
also obey the resonance condition (\ref{cond}),
while one notices again
the existence 
of a maximal horizon size for the background geometry.

\section{The Black Rings}
\label{section4}

The Tangerlini BH solution 
 \cite{Tangherlini:1963bw} 
 provides a natural higher dimensional
generalizations of the  $D=4$
Schwarzschild   solution,
possessing a horizon of spherical topology.
Nevertheless, already in 1986 
Myers and Perry argued
that
a different class of
GR solutions
with a horizon topology $S^{D-2}\times S^1$
(with $D>4$)
should  also exist \cite{Myers:1986un}.
This, indeed, bas been confirmed by the 
discovery  
in 2001
of the five dimensional black ring (BR)
by Emparan and Reall  
\cite{Emparan:2001wn,Emparan:2001wk}. 
 
The static limit of the spinning BR
(which was reported first in 
Ref. \cite{Emparan:2001wk}) 
is not fully satisfactory,
since it contains a conical singularity 
in the form of a disc ($i.e.$  a negative tension source) that sits inside the ring,
 supporting it against collapse. 
Generalizations of the  
  static vacuum BR solution    
 \cite{Emparan:2001wk}
 for more general models are known 
(see $e.g.$
\cite{Yazadjiev:2005hr,Kleihaus:2009dm})
in particular in electrovacuum
\cite{Kunduri:2004da}.
However, in  the asymptotically flat case,
 these solutions still possess conical singularities,
and 
 the only known mechanism\footnote{See, however,
the balanced BR solutions in \cite{Kleihaus:2019wck},
which are supported by a phantom scalar field.} 
to obtain a balanced configuration  
is to set the ring into
rotation \cite{Emparan:2001wn}.
In this case the centrifugal force 
 balances the ring's self-attraction for a critical horizon velocity
and a continuum of balanced solutions is found, 
with the existence of
two branches  merging
at a minimal value of the angular momentum.

There is no 
fundamental reason
to expect 
 that $all$ properties of the vacuum static BR solution, 
in particular the existence of a conical singularity, 
 hold
for any  model.  
In this Section we report on the existence of 
\textit{balanced} BR solutions in the EMgS model.  
The gauged scalar field
creates a charged environment
which provides an extra force supporting a BR against collapse.

\subsection{The ansatz and equations}
%
Given the topology difference between the horizon geometry and the
sphere at infinity,
the numerical construction of BR solutions is a
highly non-trivial numerical problem.
In this work we use a special coordinate system, 
with a single coordinate patch 
and a metric Ansatz
with four unknown functions
introduced in 
\cite{Kleihaus:2009dm,Kleihaus:2010pr},
with 
\begin{eqnarray}
\label{metric-BR}
 ds^2=-f_0(r,\theta) dt^2+f_1(r,\theta)(dr^2+r^2 d\theta^2)
+f_2(r,\theta) d\varphi_1^2
+f_3(r,\theta) d\varphi_2^2
\ .
\end{eqnarray} 
The range of $r$ is $0<r_H\leq r<\infty$, with $r_H$ the event horizon radius;
thus the $(r,\theta)$ coordinates have a rectangular boundary well suited for numerics,
the asymptotic form (\ref{metric-Mink}) being approached for $r\to  \infty$.
The scalar field and $U(1)$ potential are
still given by  (\ref{matter}),
with $V$ and $\psi$
functions of $(r,\theta)$ only.

The equations satisfied by the metric functions $f_i$ are:
\begin{eqnarray}
\label{eqs1}
&&
\nabla^2f_1 
-\frac{1}{2f_1} (\nabla f_1)^2
-\frac{1}{2} f_1
\left[
 \frac{1}{f_0f_2} (\nabla f_0)\cdot (\nabla f_2)
+\frac{1}{f_0f_3} (\nabla f_0)\cdot (\nabla f_3)
+\frac{1}{f_2f_3} (\nabla f_2)\cdot (\nabla f_3)
\right]
\\
\nonumber
&&
{~~~~~~~~~~~~~~~~~~~~~~~~}
-\frac{16 \pi G}{3}\frac{f_1}{f_0}
\left[
\frac{1}{2}(\nabla V)^2
-3f_0 (\nabla \psi)^2
+f_1(f_0 U(\psi)-3(w-g_s V)^2 \psi^2
\right]=0,
\end{eqnarray} 
\begin{eqnarray}
\nonumber
&&
\nabla^2f_2
 -\frac{1}{2f_2} (\nabla f_2)^2
+ \frac{1}{2f_0 } (\nabla f_0)\cdot (\nabla f_2)
+ \frac{1}{2f_3 } (\nabla f_2)\cdot (\nabla f_3)
+\frac{16 \pi G}{3}\frac{f_2}{f_0}
\left[
(\nabla V)^2 +2f_0 f_1 U(\psi)
\right]=0,
\\
&&
\nonumber
\nabla^2f_3
 -\frac{1}{2f_3} (\nabla f_3)^2
+ \frac{1}{2f_0 } (\nabla f_0)\cdot (\nabla f_3)
+ \frac{1}{2f_2 } (\nabla f_2)\cdot (\nabla f_3)
+\frac{16 \pi G}{3}\frac{f_3}{f_0}
\left[
(\nabla V)^2 +2f_0 f_1 U(\psi)
\right]=0,
\\
&&
\nonumber
\nabla^2f_0
 -\frac{1}{2f_0} (\nabla f_0)^2
+ \frac{1}{2f_2 } (\nabla f_0)\cdot (\nabla f_2)
+ \frac{1}{2f_3 } (\nabla f_0)\cdot (\nabla f_3)
\\
&&
\nonumber
{~~~~~~~~~~~~~~~~~~~~~~~~~~~~~~~~~~~~~~~~~~~}
-\frac{32 \pi G}{3} 
\left[
(\nabla V)^2 
+ f_1 (3(w-g_s V)^2\psi^2-f_0 U(\psi))
\right]=0,
\end{eqnarray}
while
the matter functions
$V$ and $\psi$ 
solve the
equations
\begin{eqnarray}
\label{eqs2}
&&
\nabla^2 V
-\frac{1}{2f_0} (\nabla f_0)\cdot (\nabla V)
+\frac{1}{2f_2} (\nabla f_2)\cdot (\nabla V)
+\frac{1}{2f_3} (\nabla f_3)\cdot (\nabla V)
+2g_s f_1 (w-g_s V) \psi^2=0,
\\
\nonumber
&&
\nabla^2 \psi
+\frac{1}{2f_0} (\nabla f_0)\cdot (\nabla \psi)
+\frac{1}{2f_2} (\nabla f_2)\cdot (\nabla \psi)
+\frac{1}{2f_3} (\nabla f_3)\cdot (\nabla \psi)
+\frac{f_1}{f_0}(w-g_s V)^2 \psi
-\frac{1}{2}f_1 \frac{dU}{d \psi}=0~. 
\end{eqnarray}
The nonvanishing components of the energy-momentum tensor are
\begin{eqnarray}
\nonumber
&&
T_r^{r(M)}= \frac{1}{2f_0 f_1} 
\left(  
-V_{,r}^2+\frac{1}{r^2}V_{,\theta}^2
\right), \qquad 
T_r^{r(\Psi)}=\frac{1}{f_1}
\left(
\psi_{,r}^2-\frac{1}{r^2}\psi_{,\theta}^2
\right)
+\frac{(w-g_s V)^2 \psi^2}{f_0}-U(\psi),
\\
\nonumber
&&
T_\theta^{\theta(M)}= \frac{1}{2f_0 f_1} 
\left(  
V_{,r}^2 - \frac{1}{r^2}V_{,\theta}^2
\right), \qquad 
T_\theta^{\theta(\Psi)}=\frac{1}{f_1}
\left(
-\psi_{,r}^2+\frac{1}{r^2}\psi_{,\theta}^2
\right)
+\frac{(w-g_s V)^2 \psi^2}{f_0}-U(\psi),
\\
&&
T_r^{\theta(M)}= -\frac{1}{r^2f_0 f_1} 
V_{,r} V_{,\theta}, \qquad 
T_r^{\theta(\Psi)}=\frac{2}{r^2f_1} 
\psi_{,r} \psi_{,\theta},
\\
\nonumber
&&
T_{\varphi_1}^{\varphi_1 (M)}=T_{\varphi_2}^{\varphi_2 (M)}
= \frac{1}{2f_0 f_1} (\nabla V)^2, \qquad 
T_{\varphi_1}^{\varphi_1 (\Psi)}=T_{\varphi_2}^{\varphi_2 (\Psi)}
=-\frac{1}{f_1}
(\nabla \psi)^2
+\frac{(w-g_s V)^2 \psi^2}{f_0}-U(\psi),
\\
\nonumber
&&
T_{t}^{t (M)}
= -\frac{1}{2f_0 f_1} (\nabla V)^2, \qquad 
T_{t}^{t(\Psi)} 
=-\frac{1}{f_1}
(\nabla \psi)^2
-\frac{(w-g_s V)^2 \psi^2}{f_0}-U(\psi).
\end{eqnarray}
In the above relations,  we have defined
\begin{eqnarray}
\nabla^2 A= \frac{\partial^2 A}{\partial r^2}
+\frac{1}{ r^2}\frac{\partial^2 A}{\partial \theta^2}
+\frac{1}{ r }\frac{\partial A}{\partial r }, \qquad 
(\nabla A)\cdot (\nabla B)=
\frac{\partial A}{\partial r }\frac{\partial B}{\partial r }
+\frac{1}{ r^2}
\frac{\partial A}{\partial \theta }\frac{\partial B}{\partial \theta }~.
\end{eqnarray}

\subsection{The boundary conditions  and horizon quantities }

The solutions  are found
again numerically, by
solving the
equations 
(\ref{eqs1}),
(\ref{eqs2}) 
subject to the following boundary conditions. 
We assume that as $r\to \infty$,  the Minkowski spacetime background 
(\ref{metric-Mink})
is recovered, while the scalar vanishes 
and the electrostatic potential takes a constant value. 
This implies 
\begin{equation}
\label{bc-inf}
f_0|_{r=\infty}=1,~ f_1|_{r=\infty}=1,~ \lim_{r\to \infty} \frac{f_2}{r^2}=\cos^2\theta,
\lim_{r\to \infty} \frac{f_3}{r^2}=\sin^2\theta,~ 
V|_{r=\infty}=\Phi,~~\psi|_{r=\infty}=0.
\end{equation}
At the horizon ($r=r_H>0$),
 we require 
\begin{eqnarray}
\label{bc-rh}
f_0|_{r=r_H}=0,~~
\partial_r f_1|_{r=r_H}=\partial_r f_2|_{r=r_H}=\partial_r f_3|_{r=r_H}=0,~~
V|_{r=r_H}=0,~~
\partial_r \psi |_{r=r_H}=0.
\end{eqnarray}
Turning now to the $\theta$-interval,
the boundary conditions at $\theta=\pi/2$ are
\begin{equation}
\label{bc-pi2}
\partial_\theta f_0|_{\theta=\pi/2}
=\partial_\theta f_1|_{\theta=\pi/2}
= f_2|_{\theta=\pi/2}
=\partial_\theta f_3|_{\theta=\pi/2}= 0,~
=\partial_\theta V|_{\theta=\pi/2}= 0,~
\partial_\theta  \psi|_{\theta=\pi/2}=0,~~
\end{equation}
The boundary conditions at $\theta=0$ 
are more complicated 
and imply the existence of
  a new input parameter, 
	$R>r_H$ -- the radius of the BR \cite{Kleihaus:2010pr}. 
For $r_H<r<R$, we impose 
\begin{eqnarray}
\label{bc-01}
\partial_\theta f_0|_{\theta=0}
=\partial_\theta f_1|_{\theta=0}
=f_2|_{\theta=0}
=\partial_\theta  f_3|_{\theta=0}= 0,~~
 \partial_\theta  V |_{\theta=0}=0,~~
 \partial_\theta  \psi |_{\theta=0}=0 \ ,
\end{eqnarray} 
 while the boundary conditions for $r>R$ are
\begin{eqnarray}
\label{bc-02}
\partial_\theta f_0|_{\theta=0}
=\partial_\theta f_1|_{\theta=0}
=\partial_\theta f_2|_{\theta=0}
=f_3|_{\theta=0}=0,~~
\partial_\theta \psi |_{\theta=0}=
 \partial_\theta  V |_{\theta=0}
=0.
\end{eqnarray}
This set of boundary conditions 
may look involved but they
emerge naturally 
from a study of the (electro-)vacuum limit
(see Appendix A)
and are compatible with the requirement that
the solutions describe
 asymptotically flat BRs.
For example, 
 at $\theta=0$
for some interval $r_H\leq r<R$, 
we have  the same conditions as for $\theta=\pi/2$,
with a nonzero metric function 
$g_{\varphi_1 \varphi_1}$
and a vanishing 
$g_{\varphi_2 \varphi_2}$,
while
 the boundary conditions for $r> R$
are compatible with
the asymptotic behaviour
 $g_{\varphi_1 \varphi_1} \sim \cos^2\theta$, 
$g_{\varphi_2 \varphi_2}  \sim \sin^2 \theta$.

Let us mention that apart from (\ref{bc-inf})-(\ref{bc-02}), the solutions 
should also satisfy a
number of extra-conditions
(like the constancy of the surface gravity
or the absence of conical singularities at $\theta=\pi/2$
and $\theta=0$, $r>R$),
which originate mainly from the constraint equations.
However, these extra-conditions
are not imposed in the numerics; rather, they are used to verify the accuracy of the results.

The  metric of a spatial cross-section of the horizon is
\begin{eqnarray}
\label{horizon-metric}
d\sigma^2=f_1(r_H,\theta)r^2_H d\theta^2+f_2(r_H,\theta)d\varphi_1^2+f_3(r_H,\theta)d\varphi_2^2.
\end{eqnarray} 
The orbits of $\varphi_1$ shrink 
 to zero at $\theta=0$ and $\theta=\pi/2$,
 while 
the length of $S^1$-circle does not vanish anywhere, such that 
the topology of the horizon 
is $S^2\times S^1$ 
(in fact,   $f_2(r_H,\theta)\sim \sin^2 2\theta$
while $f_1(r_H,\theta)$ and $f_3(r_H,\theta)$ are strictly positive and finite functions).
The
event horizon area and the Hawking temperature  are given by
\begin{eqnarray}
\label{AH}
A_H=4\pi^2 r_H \int_{0}^{\pi/2} d\theta \sqrt{f_1 f_2f_3 }\bigg|_{r=r_H}~,~~
T_H= 
\frac{1}{2 \pi }\lim_{r\to r_H}\sqrt{\frac{f_0}{(r-r_H)^2f_1}} ~.
\end{eqnarray}

As expected, 
the generic configurations possess a conical singularity.  
The strength of this singularity is measured by the parameter
\begin{eqnarray}
\label{delta}
\delta= 2\pi \bigg(1-\lim_{\theta\to 0}\frac{f_2}{\theta^2 r^2 f_1 }\bigg) \neq 0~,
\end{eqnarray}
for $\theta=0$ and $r_H<r<R$.
This can be interpreted as a disk
preventing the collapse of the configurations.
Despite the presence of a conical singularity,
the solutions still
admit a thermodynamical description.
Moreover,  
the entropy of these {\it unbalanced} BRs is still one quarter of the 
event horizon area, 
with the parameter $\delta$
entering the first law of thermodynamics as a pressure term,
see $e.g.$
the discussion in \cite{Herdeiro:2009vd,Herdeiro:2010aq}.

Finally, we mention that the 
boundary conditions 
 (\ref{bc-inf})-(\ref{bc-02})
are compatible with  an approximate expansion of 
the solutions at the boundaries.
Then this study shows that
both the resonance 
and the bound state
conditions
(\ref{cond}),
(\ref{bound}) 
hold also for a BR. 
For example,
as with BHs and BSs,  
we assume the existence 
of a power series expansion of the solutions in $(r-r_H)$ 
(with $f_0\sim (r-r_H)^2$
and $f_{1,2,3}(r_H)$ nonzero) close to the horizon.
Then the condition (\ref{cond})
naturally emerges, albeit this time the 
series coefficients are
 $\theta$-dependent.

\subsection{The numerical approach }
 
Following the approach in
\cite{Kleihaus:2009dm,Kleihaus:2010pr}
(see also
the related work
\cite{Kleihaus:2010hd,Kleihaus:2012xh,Kleihaus:2014pha})
we define
 \begin{eqnarray}
 \label{bgf2} 
 {f}_i= f_i^{(0)}e^{{F}_i} \ ,
\end{eqnarray}
with the {\it background functions}
 $ {f}_i^{(0)}$, 
which
are those of the vacuum 
static 
BR, as given 
in Appendix A, eqs. (\ref{BR-vacuum}).
In this approach, the coordinate singularities
are essentially subtracted, 
while 
automatically
imposing at the same time the
BR
event horizon topology
together with
 the required boundary behaviour of the metric functions. 
 
The new  functions $F_i$  
encode the effects of ``deforming''
the vacuum BR by the matter fields.
They
are smooth and finite everywhere, 
such that they do not 
lead to the occurrence of new zeros of the
metric  functions $f_i$. 

The numerics is done in terms of the functions 
$F_i$ which enter (\ref{bgf2}), subject to
 Newman boundary conditions on all boundaries except at infinity,
where we impose $F_i=0$
 (these boundary conditions 
follow  from 
(\ref{bc-inf})-(\ref{bc-01}),
together with (\ref{BR-vacuum})). 
Also, instead of $r$,
in the numerics we use a new
compactified radial coordinate
$x=(r-r_H)/(c+r)$, with $c$ a properly chosen constant.
The equations for $(F_i; V,\psi)$
(which result directly from (\ref{eqs1}), (\ref{eqs2}))
 are solved 
by employing a finite difference solver \cite{schoen},
which uses  a Newton-Raphson method
(we  use an order six for the discretization of derivatives).
This professional software provides an error estimate for
each unknown function,  which is typically lower than
than $10^{-3}$.

\subsection{The results }

For a BR, apart from the theory parameters
$(\alpha,\beta,e)$
and the chemical potential $\Phi$,
there are two more input parameters 
$(r_H,R)$.
Although $(r_H,R)$ have no invariant meaning,
they still provide 
a rough measure for the radii of the 
$S^2$  and $S^{1}$ parts in the horizon metric (\ref{horizon-metric}).

\begin{figure}[ht!]
\begin{center} 
\includegraphics[height=.34\textwidth, angle =0 ]{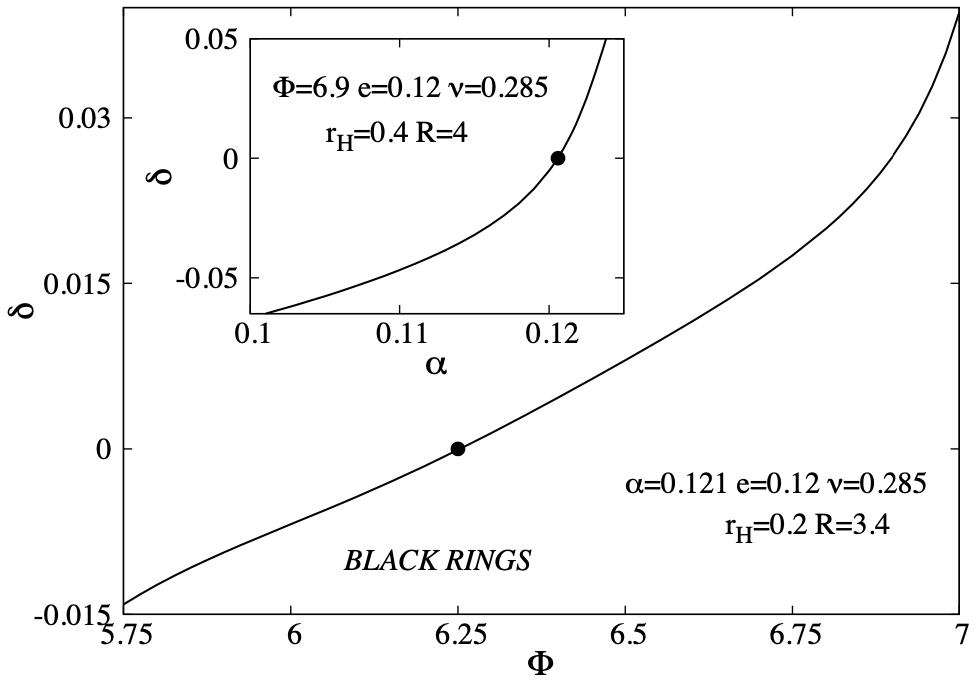}
\includegraphics[height=.34\textwidth, angle =0 ]{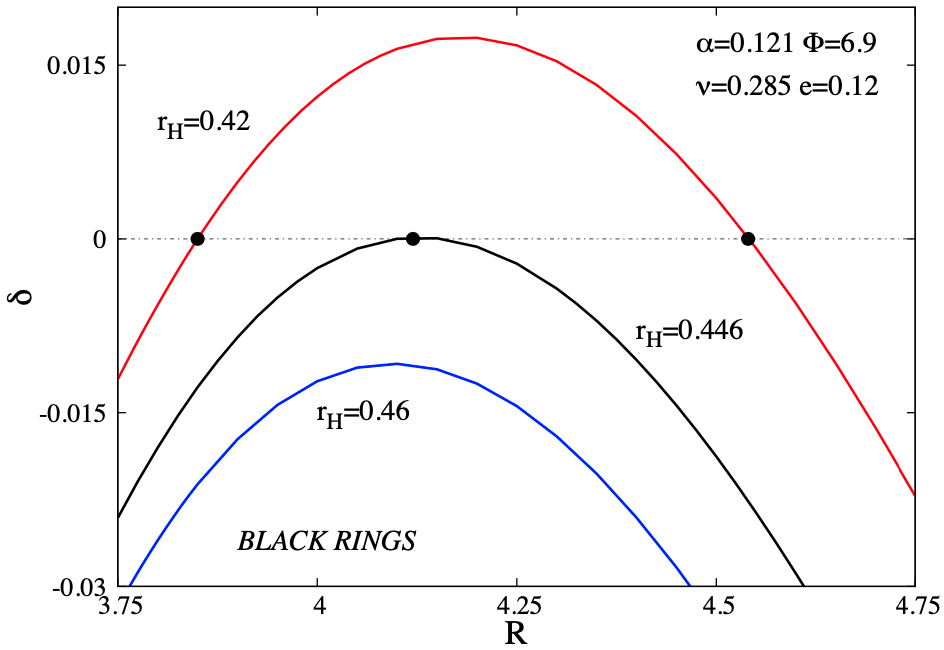}
\end{center}
\caption{
The conical excess/deficit 
$\delta$  
of a static BR in EMgS field theory
is shown as a function of the
input parameter $\alpha$ which measures the strength of gravity
(the inset)
and  of the electrostatic potential $\Phi$ 
 (left panel),
and as function of the ring's radius $R$   (right panel). 
}
\label{BR1}
\end{figure}

The reported solutions are regular on and outside the horizon and
show no sign of a singular behaviour, 
apart from the generic existence 
of conical singularities.
However, 
balanced BRs exist as well, requiring a fine-tunning of the input parameters.
They are constructed as follows.  
The starting point are the 
solutions in the probe limit, 
$i.e.$
with $\alpha=0$, some values of 
$(\beta,e;\Phi)$
and a vacuum BR background with parameters
$(r_H,R)$.
Then the value of the coupling constant
$\alpha$ is increased in small steps.
As seen in Figure \ref{BR1} (inset),
the (absolute) value of 
the conical excess $\delta$
decreases as $\alpha$
is increased.
Therefore, for a BR set with fixed
horizon and ring radii $(r_H,R)$, 
a balanced configuration
(marked with a black dot in that figure)
is achieved for a critical value of $\alpha$.
Further increasing $\alpha$
results in configurations with a 
conical excess $\delta>0$.
Balanced BR exist as well for a critical value of 
the chemical potential $\Phi$
($i.e.$ a strong enough electric field)
when keeping fixed the other input parameters, 
see Figure \ref{BR1} (left panel).

\begin{figure}[ht!]
\begin{center}
\includegraphics[height=.34\textwidth, angle =0 ]{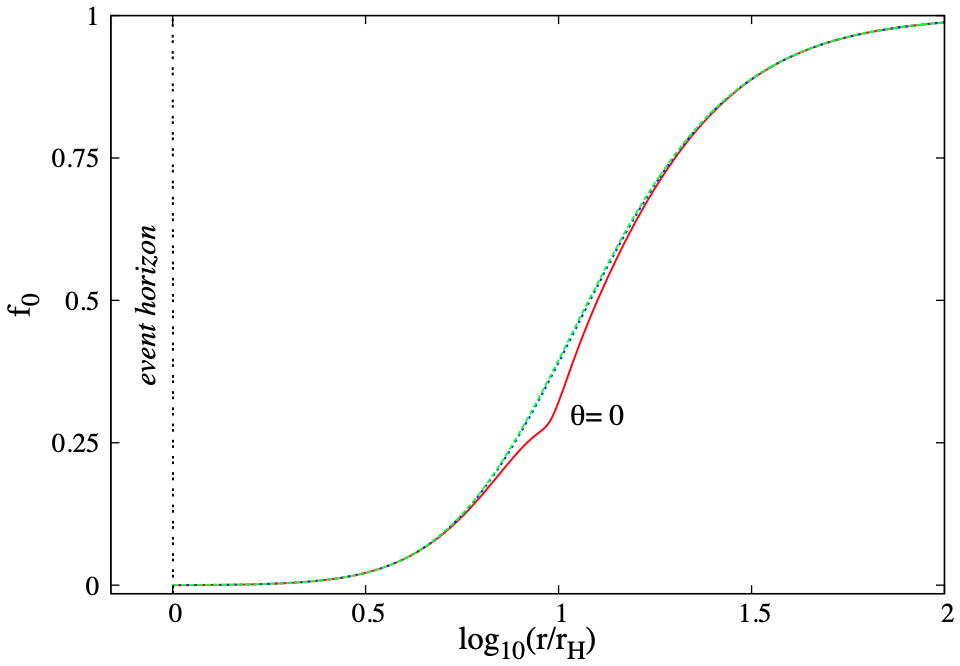} 
\includegraphics[height=.34\textwidth, angle =0 ]{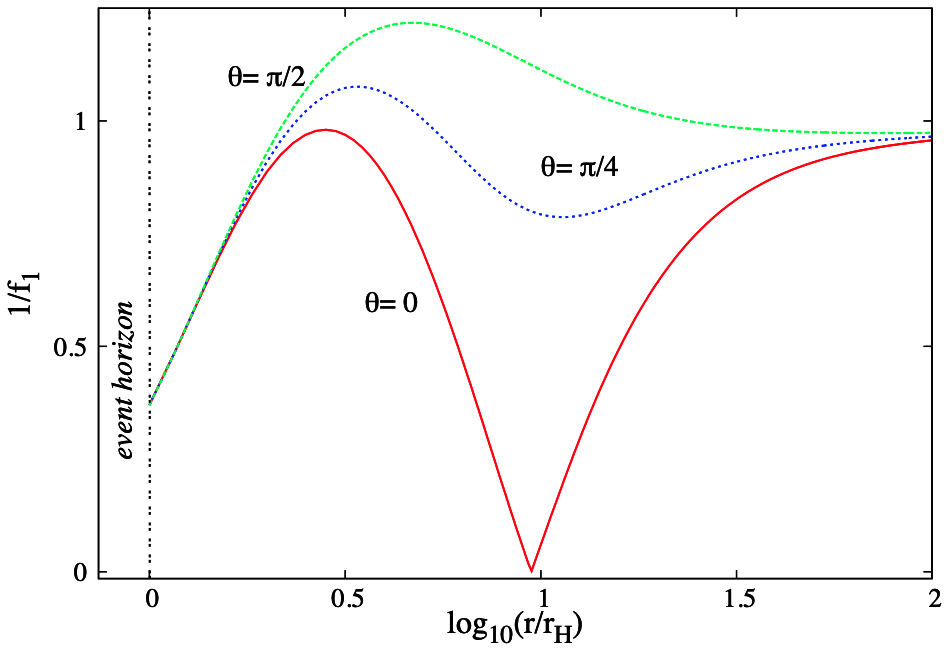}
\includegraphics[height=.33\textwidth, angle =0 ]{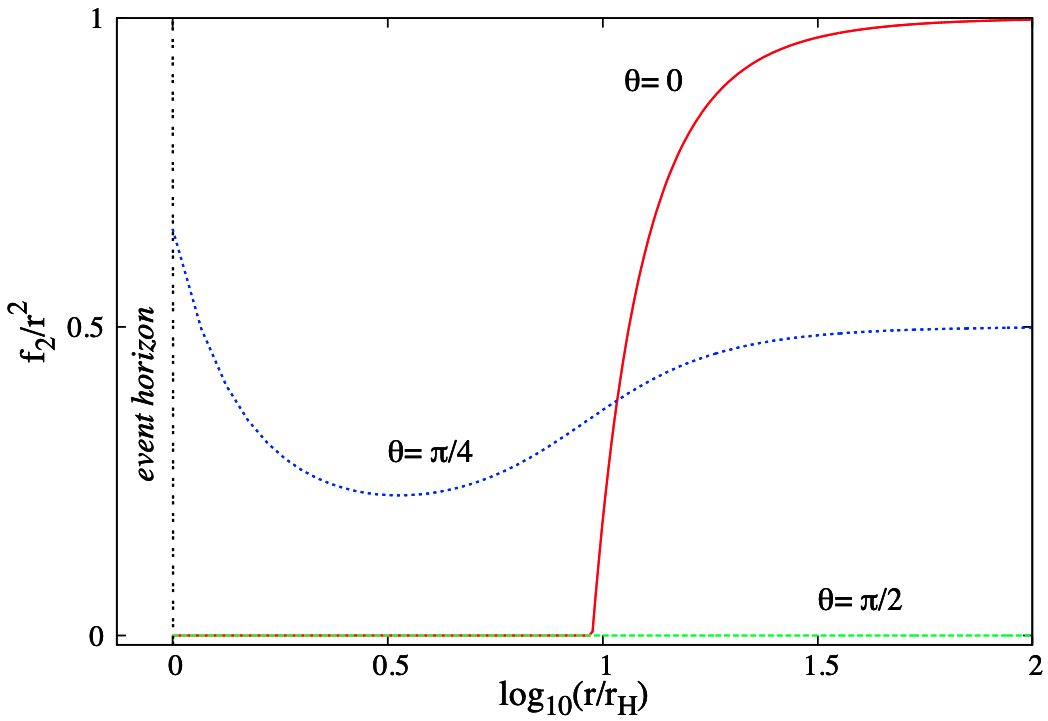} 
\includegraphics[height=.33\textwidth, angle =0 ]{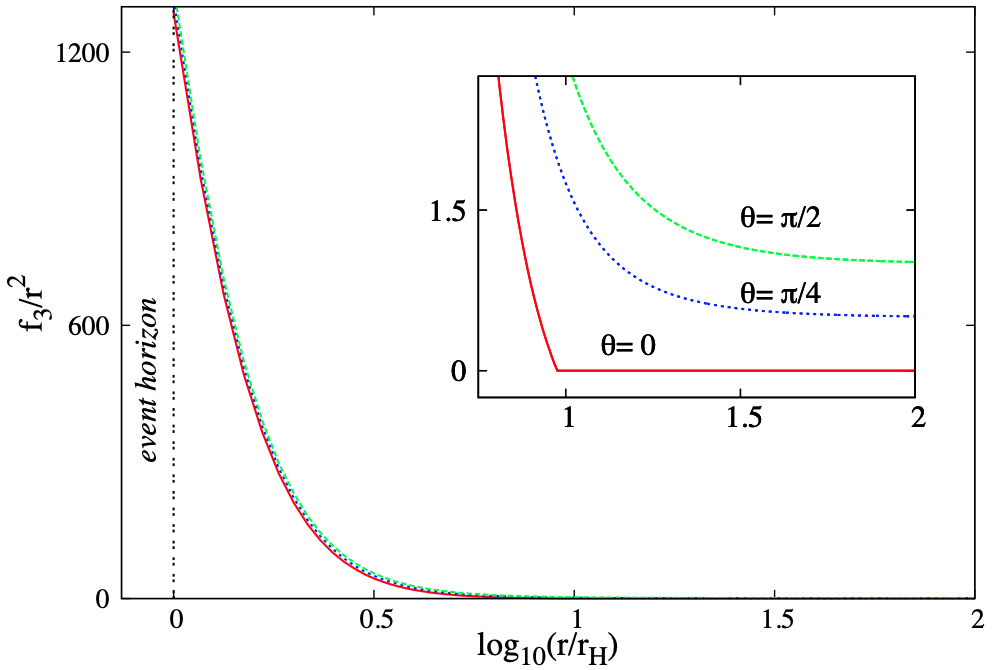}
\includegraphics[height=.34\textwidth, angle =0 ]{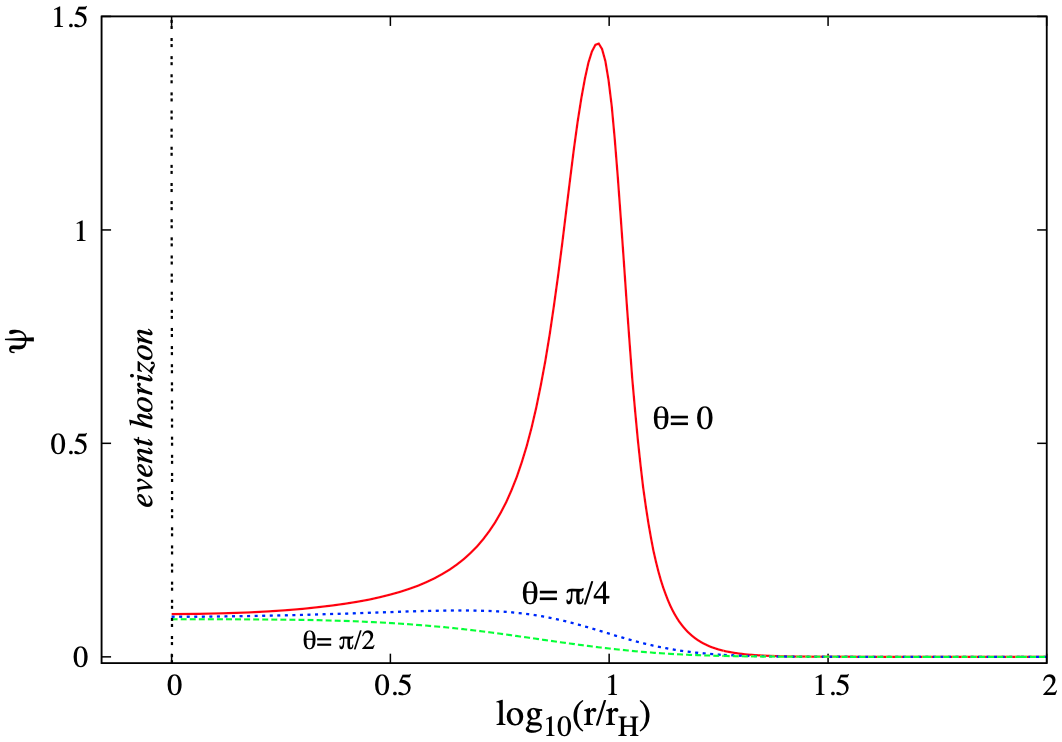} 
\includegraphics[height=.34\textwidth, angle =0 ]{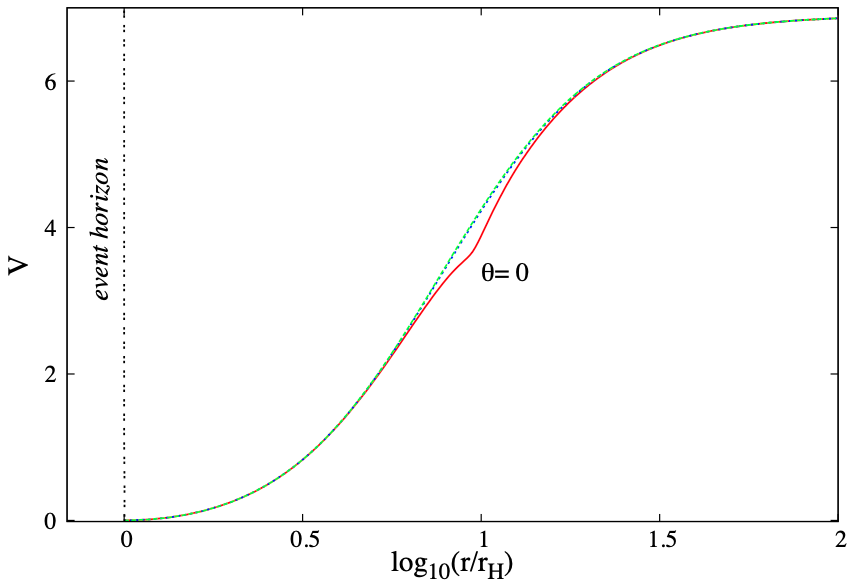}
\end{center}
\caption{
Profile functions of a typical static
 BR solution with gauged scalar hair 
are shown  for three values of the
angular coordinate $\theta$.
The input parameters are 
$r_H=0.4$,
$R=3.781$
and
$\alpha=0.121$,
$\nu=0.285$,
$\Phi=6.9$,
$e=0.12$. 
}
\label{BR-profile}
\end{figure}

The profile of a typical balanced solution
$(\delta=0)$
is shown in Figure \ref{BR-profile}.
As one can see, the functions $f_i$
display a complicated behaviour, which, however,
 is compatible with the
imposed boundary conditions, being essentially
fixed by the  $ {f}_i^{(0)}$ contribution in (\ref{bgf2}).
Also, while the electric potential has a small angular dependence,
this is not the case for the scalar field,
which possesses a maximum at 
$r=R$,
$\theta=0$.
 We remark there are no nodes, with $\psi(r,\theta)>0$ not crossing the axis.

In the context of this work, we are interested in 
solutions with a fixed 
$\Phi$ and fixed coupling constants  
$(\alpha>0,\beta, e)$.
Then, 
when varying  the size of the ring (via the input parameter $R$)
for fixed $r_H$,
the conical deficit/excess $\delta$
varies as well,
which may result in the 
possible
existence of  balanced configurations, $\delta=0$.
In Figure \ref{BR1} (right panel) 
the value of $\delta$ 
is shown
as a function of $R$
for several value of the horizon radius $r_H$.
One can see that for
some range
$r_H<r_H^{(\rm max)}$
the condition 
$\delta=0$
is satisfied by two different
BR solutions (marked with dots in Figure \ref{BR1}).
The value of  
$r_H^{(\rm max)}$, however, 
depends on the choice of other input parameters.

\begin{figure}[ht!]
\begin{center} 
\includegraphics[height=.34\textwidth, angle =0 ]{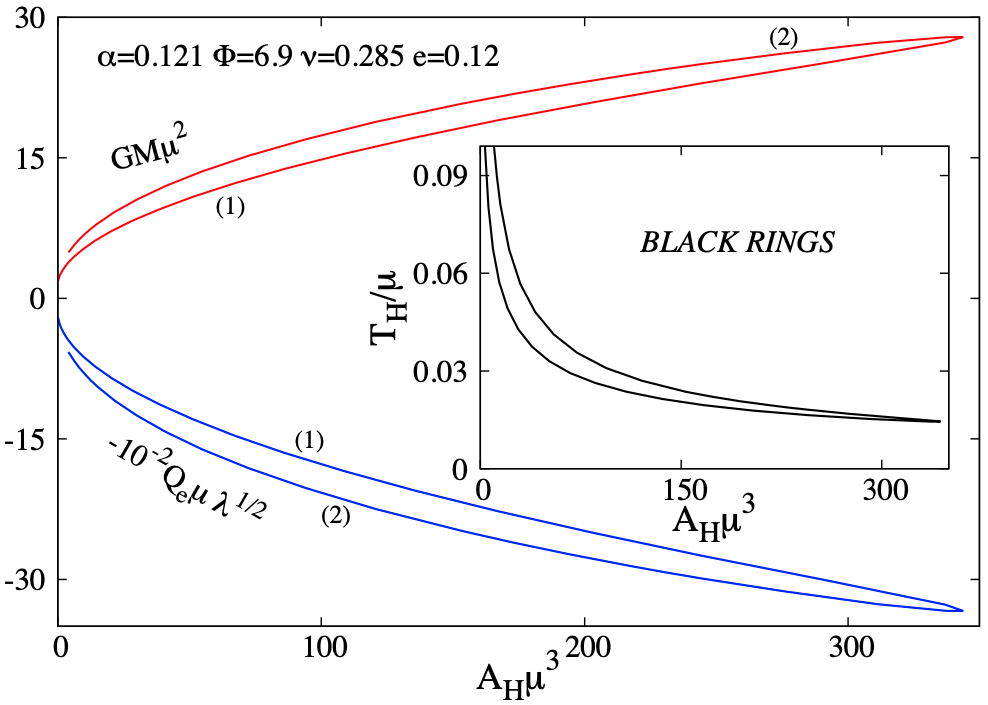}
\includegraphics[height=.34\textwidth, angle =0 ]{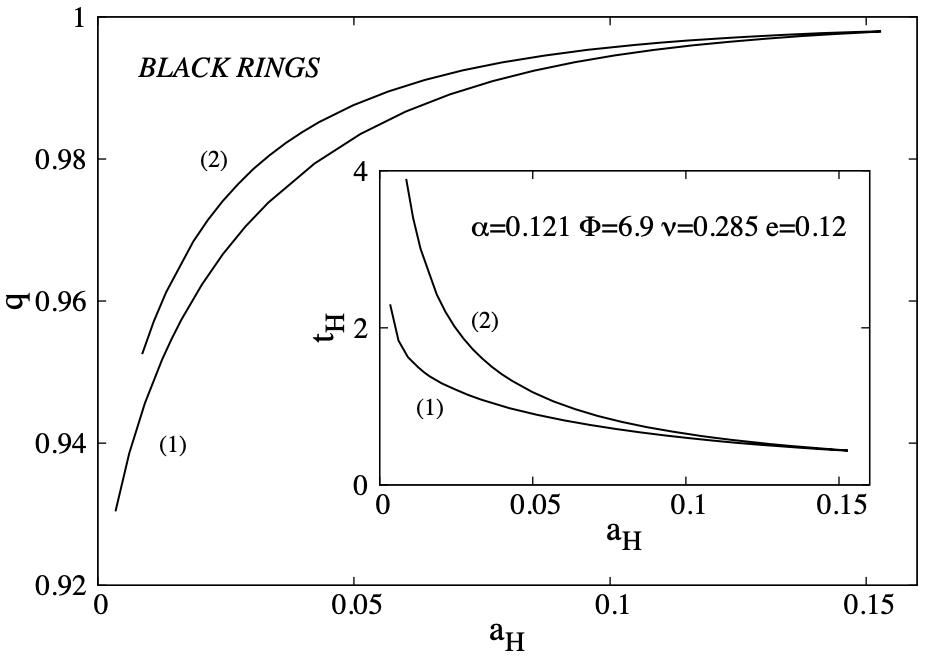}
\end{center}
\caption{
The mass, electric charge and Hawking temperature of solutions
are shown as a function of event horizon area
for balanced BR solutions.
}
\label{BR2}
\end{figure}

Some basic features of the solutions with
fixed $\Phi$ can already be deduced from 
Figure \ref{BR1} (right panel).
First, all balanced BRs
exist up to a maximal horizon size.
That is, no arbitrarily large 
balanced
BRs 
were found for the cases investigated so far. 
Second, 
one notices
again 
 the existence of two branches of solutions
merging for some maximal horizon size, 
see
Figure \ref{BR2}. 

Although further work is necessary, 
the existing results suggest a strong analogy
with the picture found for BHs and BSs.
A branch of fundamental solutions
(label (1) in Figure  \ref{BR2})
 emerges from 
the (spherically symmetric)
gravitating gauged $Q$-ball
when adding a small horizon with ring topology. 
As seen in  Figure \ref{BR2},  
this set stops to exist for a maximal value of the horizon area,  
with a backbending and the occurrence of a secondary branch. 
The limiting solution
of this branch is unclear, 
even though
the situation  
appears to be similar to that found in \cite{Herdeiro:2020xmb}
for $D=4$ BHs.
That is, the secondary branch stops to exist for a  
nonzero $A_H$, where
the numerical results become unreliable,
with large violations of the constraint equations. 

We expect to find a different picture for other values of
the electrostatic potential,
with the  existence 
of two branches interpolating between different  
(spherically symmetric) 
gauged $Q$-balls.
In any case, the presence of scalar hair leads to a very different
picture as compared to that found for
BRs in electrovacuum, as 
$e.g.$ seen when comparing  Figures
\ref{BR2}
(right panel)
and
\ref{EMBR}. 

Finally, let us mention that, 
in the (electro)vacuum case, the
spherically symmetric BHs are recovered as the limit 
$r_H\to R$ of the BR solution.
However,
somehow unexpected,
so far
we did not find any indication for the existence of 
such a smooth limit with scalar hair.
That is, all EMgS solutions with $r_H$ nonzero and close to $R$
possess conical singularities.

\section{Summary and overview}
\label{section5}

In this paper we have reported the first construction 
of higher dimensional ($D > 4$)  
solutions with charged scalar hair in the literature.
Three different types of static solutions have been 
considered, corresponding to black holes (BHs),
black strings (BSs) and black rings (BRs).
One of the conclusions of our study
is the confirmation that the properties of 
known $D=4$ BHs 
\cite{Hong:2019mcj,Herdeiro:2020xmb,Hong:2020miv}
are generic, 
their existence being anchored in the resonance
condition (\ref{cond}).
As a byproduct of our study,
we report on the existence of
static BRs which are free of singularities,
on and outside the event horizon. 

As possible avenues for future research, we start
with a number of 
open issues which apply also for 
the known four dimensional solutions.
In our opinion, 
a main question is to clarify why no hairy solutions
could be found in the absence of
scalar field self-interaction,
$i.e.$ with $\lambda=\nu=0$
in the scalar field potential (\ref{potential}).
Moreover, it would be interesting to investigate
the issue of linear stability of the $D=4,5$
hairy BHs (or for the $D=5$ BSs).

Turning to more general models,
we predict the existence of similar solutions for a gauged
Proca field.
So far only
$D=4$ solutions without an event horizon  have been discussed
in the literature, see $e.g.$ \cite{Garcia:2016ldc}.

The status of the EMgS solutions with more general asymptotics
is also worth studying.
It is interesting to remark 
that
such solutions
are known to exist for AdS asymptotics, providing the gravity duals of $s-wave$
superconductors \cite{Gubser:2008px}.
The main difference $w.r.t.$ the asymptotically flat case is that
the AdS solutions emerge as perturbations around a RN-AdS background.
Also, the nonlinearities of the scalar field potential
play no important role in this context.
It would certainly be interesting to look for similar solutions 
with a positive cosmological constant.
 
There are also a number of
issues specific to higher dimensions.
 First, it would be interesting to
 clarify the generality of the
resonance mechanism:
dos it work for 
any horizon topology and any dimension $D\geq 4$?
Furthermore, 
for the same matter content
and $D=5$,
one could
 consider other type of solutions
in Kaluza-Klein theory, such as 
caged BHs 
\cite{Kudoh:2004hs,Kalisch:2018efd}
and squashed BHs
\cite{Ishihara:2005dp}. 

Finally, let us mention that
the (electro)-vacuum solutions with
  Kaluza-Klein asymptotics possess
the (gravitational) Gregory-Laflamme instability
\cite{Gregory:1993vy},
with the additional existence 
of another set of solutions
depending of the compact extra-coordinate, which is $z$
for the background metric (\ref{metric-KK}),   
$i.e.$ nonuniform BSs
\cite{Gubser:2001ac,Wiseman:2002zc,Kleihaus:2006ee}. 
On general grounds, we expect the
existence of similar solutions in the considered
EMgS model.

\section*{Acknowlegements}

This work is supported by the Center for Research and Development in Mathematics and Applications (CIDMA) through the Portuguese Foundation for Science and Technology (FCT - Funda\c{c}\~{a}o para a Ci\^{e}ncia e a Tecnologia), references UIDB/04106/2020, UIDP/04106/2020.  We acknowledge support from the FCT funded research grants PTDC/FIS-OUT/28407/2017, \\ CERN/FISPAR/0027/2019, PTDC/FIS-AST/3041/2020 and CERN/FIS-PAR/0024/2021. This work has further been supported by the European Union's Horizon 2020 research and innovation (RISE) programme H2020-MSCA-RISE-2017 Grant No. FunFiCO-777740.

 \appendix
\section{The hairless limit: 
Black holes, strings and rings in Einstein-Maxwell theory }
\setcounter{equation}{0}
\renewcommand{\theequation}{A.\arabic{equation}}

\subsection{ The  Reissner-Nordstr\"om black hole}
This is the natural generalizations of the well known four dimensional solution,
possessing the same basic properties.

\begin{figure}[ht!]
\begin{center} 
\includegraphics[height=.34\textwidth, angle =0 ]{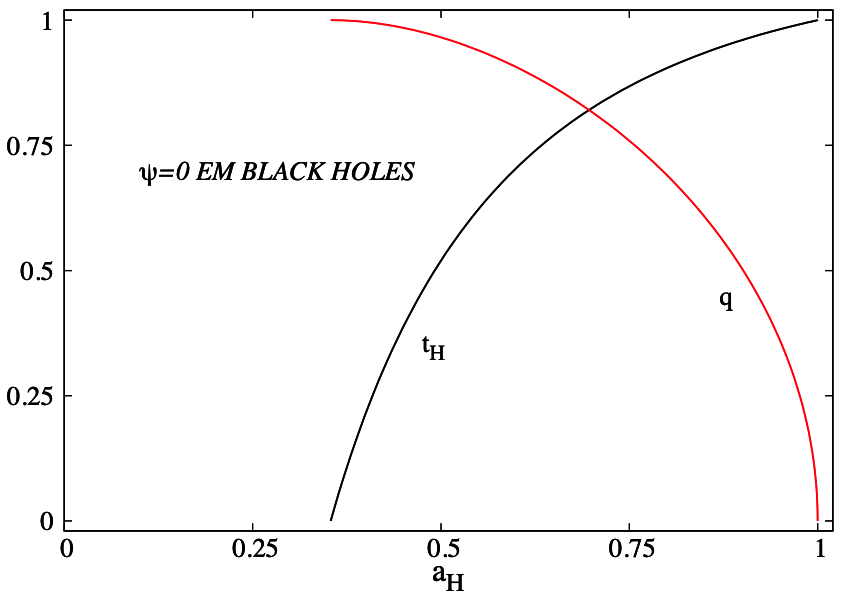}
\includegraphics[height=.34\textwidth, angle =0 ]{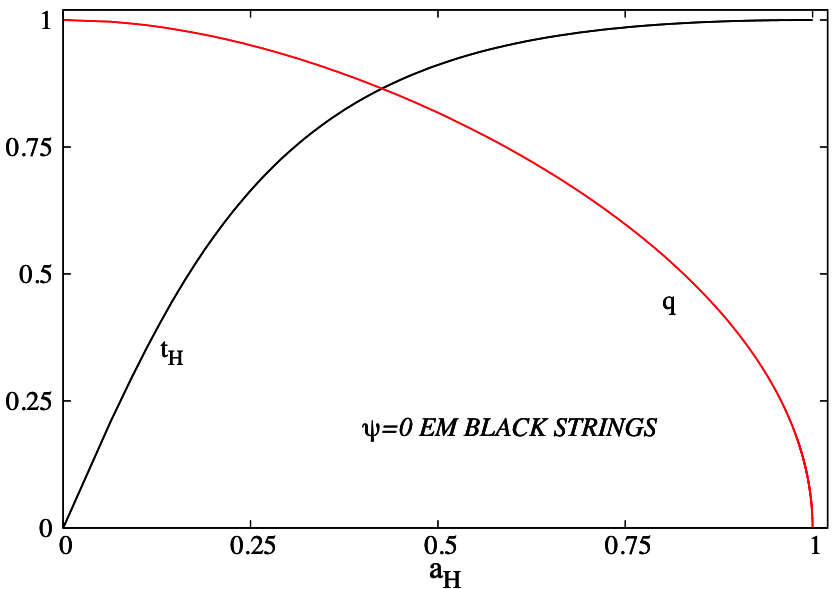}
\end{center}
\caption{
The reduced charge and Hawking temperature are shown as a function of
reduced event horizon area
for the BH
and the BS
 solutions in Einstein-Maxwell theory.
}
\label{RNBHBS}
\end{figure}

The line element and the electrostatic potential of the $D=5$ solution reads
\begin{equation}
ds^2=-N(r)dt^2+\frac{dr^2}{N(r)}+r^2 d\Omega_3^2, \quad 
N(r) = 1-\frac{8G}{3\pi} \frac{M}{r^2}+\frac{G Q_e^2}{3\pi^3}\frac{1}{r^4},
\quad 
V(r)=\frac{Q_e}{4\pi^2 r_H^2}-\frac{1}{4\pi^2}\frac{Q_e}{r^2}.~~{~~~}
\end{equation}
Instead of $M$, 
it is convenient to work with 
$r_H$,
the event horizon radius
(with $N(r_H)=0$).
Then the expressions of various quantities of interest are
\begin{equation}
M=\frac{3\pi r_H^2}{8 G}\bigg(1+\frac{G Q_e^2}{3\pi^3 r_H^4}\bigg),~~
A_H=2\pi^2 r_H^3,~~
T_H=\frac{1}{2\pi r_H}\bigg(1-\frac{G Q_e^2}{3\pi^3 r_H^4}\bigg),~~
\Phi=\frac{Q_e}{4\pi^2 r_H^2}.
\end{equation}

The reduced quantities 
 $a_H$ and $t_H$
as defined by
(\ref{scale-BHR}),
can 
be expressed as a function of the
scaled charge $q$,
with
\begin{eqnarray}
a_H=\frac{1}{2\sqrt{2}}(1+\sqrt{1-q^2})^{3/2}, \qquad 
t_H= \frac{2\sqrt{2}\sqrt{1-q^2}}{(1+\sqrt{1-q^2})^{3/2}}.
\end{eqnarray} 
As one can see in Figure \ref{RNBHBS} (left panel),  
 $a_H$ varies between one  -- the Schwarzschild-Tangherlini limit, with
$q=0$, 
$t_H=1$ -- ,
and $1/(2\sqrt{2})$ --
the extremal limit,
with
 $q=1$
and 
$t_H=0$.

\subsection{ The charged black string}

The line element and the electrostatic potential of
the BS are given by
\begin{equation}
ds^2=-\frac{F(r)}{\Omega(r)^2} dt^2+
\Omega(r)
\left[
\frac{dr^2}{F(r)}+r^2 d\Omega_2^2+dz^2
\right]
,
\qquad
V(r)=\frac{Q}{2\pi r_H}\frac{1}{1+\sqrt{1+\frac{4G Q^2}{3\pi r_H^2}}}\frac{F(r)}{\Omega(r)} \ ,
\end{equation}
where
\begin{eqnarray}
F(r)=1-\frac{r_H}{r},~~
\Omega(r)=1+\frac{r_H}{2r}
\bigg( \sqrt{1+\frac{4G Q^2}{3\pi r_H^2}}-1\bigg)~.
\end{eqnarray}
This solution has two input parameters 
$r_H$ and $Q$,
with the following expression for various quantities of interest
\begin{eqnarray}
&&
M=\frac{Lr_H}{8G}\bigg(1+3  \sqrt{1+\frac{4G Q^2}{3\pi r_H^2}}\bigg), \quad
{\cal T}=\frac{r_H}{4 G}, \quad
Q_e =L Q, \quad 
T_H=\frac{1}{\sqrt{2}\pi r_H}\frac{1}{\bigg(1+\frac{4G Q^2}{3\pi r_H^2}\bigg)^{3/2}},
\nonumber
\\
&&
A_H= \sqrt{2}\pi L r_H^2
\left(
1+\sqrt{1+\frac{4G Q^2}{3\pi r_H^2}}
\right)^{3/2},~~
\Phi= \frac{Q}{2\pi r_H}\frac{1}{1+\sqrt{1+\frac{4G Q^2}{3\pi r_H^2}}}~.
\end{eqnarray}   
From (\ref{scale-BS}),
we get the following expression
of the scaled area, electric charge and temperature:  
\begin{eqnarray}
a_H=\frac{4\sqrt{2x}(x+\sqrt{1+x^2})^{3/2}}{(x+3\sqrt{1+x^2})^2}, \quad
q=\frac{1}{\sqrt{1+x^2}+\frac{x}{3}}, \quad 
t_H=\frac{(x+3\sqrt{1+x^2})\sqrt{x}}{\sqrt{2}(x+\sqrt{1+x^2})^{3/2}},
\end{eqnarray}	
with $0\leq x<\infty$
an arbitrary parameter. 
The extremal BS limit is singular in this case, being approached for
$x=0$ with $a_H=t_H=0$ and $q=1$.
A $x\to \infty$, the Schwarzschild BS is recovered,
with $a_H=t_H=1$
and $q=0$, see Figure \ref{RNBHBS} (right panel).

\subsection{ The static charged black ring} 

This solution has being reported in \cite{Kunduri:2004da}
in the Einstein-Maxwell-dilaton model
and
for a ring coordinate system.
Here we discuss its basic properties
for the coordinate system employed 
in this work and a vanishing dilaton field.  

Its line element and the electic potential reads
\begin{eqnarray}
&&
\label{BR-EM}
ds^2=-\frac{f_0^{(0)} (r,\theta) }{\Omega(r,\theta)^2} dt^2+
\Omega(r,\theta)
\left[
 f_1^{(0)}(r,\theta) (dr^2+r^2 d \theta^2)
+f_2^{(0)}(r,\theta)  d\varphi_1^2
+f_3^{(0)}(r,\theta)  d\varphi_2^2   
\right],
\nonumber
\\
&&
V(r,\theta)=\sqrt{\frac{3}{16 \pi G} } 
\frac{f_0^{(0)}(r,\theta) \tanh \gamma }{\Omega(r,\theta)},
~~~{\rm where}~~
\Omega(r,\theta)=\cosh^2 \gamma-\sinh^2 \gamma f_0^{(0)} (r,\theta) ,
\end{eqnarray}
with
$\gamma$  
an arbitrary real parameter.

The functions $f_i^{(0)}(r,\theta)$
$(i=0,\dots,3)$
corresponds to those which enter
 (static) Emparan-Reall vacuum solution,
with the following expression 
\cite{Kleihaus:2014pha}:
\begin{eqnarray}
\label{BR-vacuum}
&&
 f_0^{(0)}  
=  \frac{\big(1-\frac{r_H^2}{r^2}\big)^2}{\big(1+\frac{r_H^2}{r^2}\big)^2},~~
  f_1^{(0)}  
= \frac{\big(1+\frac{r_H^2}{r^2}\big)^2}{\big(1+\frac{r_H^2}{R^2} \big)^2 P }
\bigg[
\bigg(1+\frac{r_H^4}{r^4}\bigg) \bigg(1+\frac{r_H^4}{R^4}\bigg) -\frac{4r_H^4}{r^2 R^2}\cos 2\theta +\frac{2r_H^2}{R^2}P 
\bigg] , \ \ \ \ \  \ \ \ \ \ 
\\
 &&
f_2^{(0)}=\frac{1}{ f_3^{(0)}}r^4\bigg(1+\frac{r_H^2}{r^2}\bigg)^4 \sin^2 \theta \cos^2 \theta,
~~
 f_3^{(0)}=
\frac{r^2}{2}
\bigg[
P+ \frac{R^2}{r^2}
\bigg(
1+\frac{r_H^4}{R^4}-\frac{r_H^2}{R^2}\bigg(\frac{r^2}{r_H^2}+\frac{r_H^2}{r^2}\bigg)\cos 2 \theta
\bigg)  
\bigg]
 \ ,
\nonumber
\end{eqnarray}
where we define
\begin{equation}
P= \frac{r^2}{2}\left[
\left(1+\left(\frac{R}{r}\right)^4
         -2 \cos 2\theta \left(\frac{R}{r}\right)^2\right)
\left(1+\left(\frac{r_H^2}{r R}\right)^4
         -2\cos 2\theta  \left(\frac{r_H^2}{r R}\right)^2\right)	 
\right]^{1/2} \ ,
\nonumber
\end{equation}
%
\begin{figure}[ht!]
\begin{center} 
\includegraphics[height=.34\textwidth, angle =0 ]{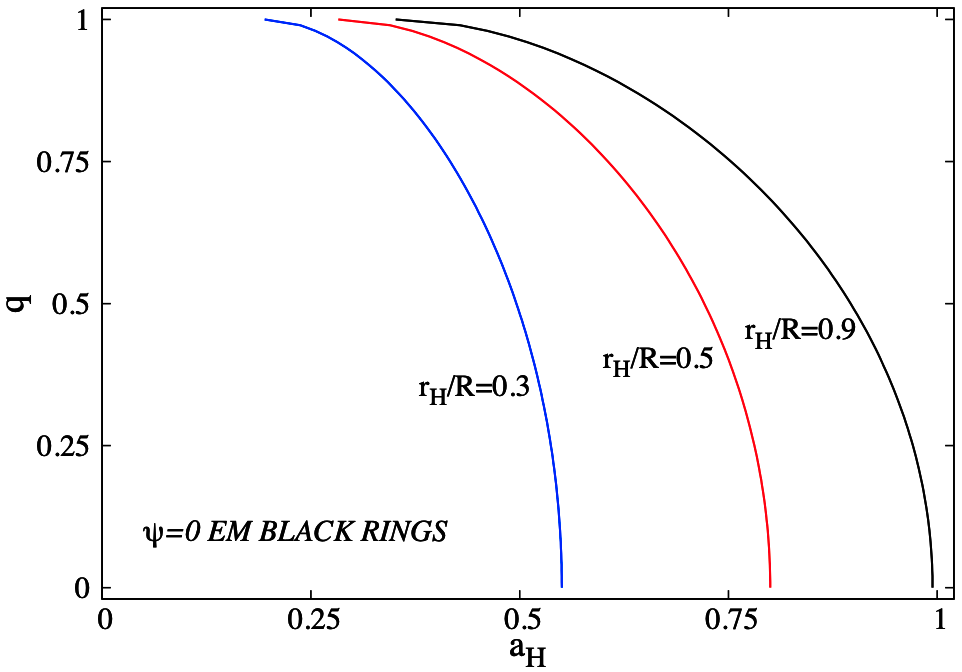}
\includegraphics[height=.34\textwidth, angle =0 ]{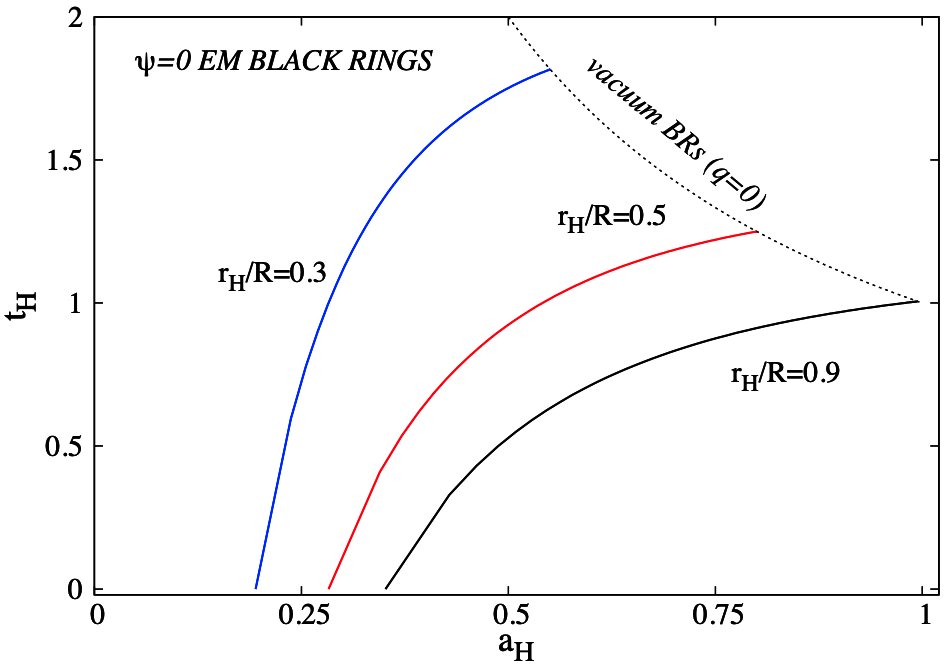}
\end{center}
\caption{
Same as Figure~\ref{RNBHBS}
for charged BRs.
}
\label{EMBR}
\end{figure}
with  $r_H$ the event horizon radius
and 
$R>r_H$   the radius of the ring.
The vacuum static BR is recovered for 
$\gamma=0.$
A discussion of the 
 properties this limiting solution for the above parametrization 
(including
the correspondence with the
better known
Weyl coordinates),
 can be found in \cite{Kleihaus:2009dm,Kleihaus:2010pr}.
Here we focused on its charged generalization.

Although the geometry (\ref{BR-EM})
is asymptotically 
flat,
it contains a conical singularity for
 $\theta=0$ and $r_H\leq r<R$,
with the following expression of
$\delta$, as resulting from (\ref{delta}):
\begin{eqnarray}
\delta= - \frac{4\pi r_H^2}{(R^2-r_H^2)}<0~.
\end{eqnarray}

Also,
the BH limit of the solutions is recovered as $r_H\to R$;
the BS limit is more subtle, being
approached 
as $R\to \infty$,
together with a suitable scaling of 
$r$,
$r_H$
and 
$\gamma$.

The expression of various quantities of interest are
\begin{eqnarray}
&&
M=\frac{3\pi r_H^2 \cosh (2\gamma)}{2G},~~
Q_e=\frac{2\sqrt{3}\pi^{3/2}r_H^2\sinh(2\gamma)}{\sqrt{G}},~~
T_H=\frac{1+\frac{R^2}{r_H^2}}{8\pi R}\frac{1}{\cosh^3 \gamma},
\\
&&
A_H=\frac{32 \pi^2 R r_H^4 \cosh^3 \gamma}{R^2+r_H^2},~~
\Phi=\sqrt{\frac{3}{16 \pi G}}\tanh \gamma~.
\end{eqnarray}
One can verify that the Smarr relation
(\ref{Smarr})
is still satisfied, but not 
the 1st law
(\ref{1st}),
which requires to introduce an extra term  
associated with the conical singularity
\cite{Herdeiro:2009vd,Herdeiro:2010aq}.

Concerning the reduced quantities (\ref{scale-BHR}),
both $a_H$ and $t_H$
can 
be expressed as a function of the
scaled charge $q$,
with
\begin{equation}
a_H=\frac{x}{\sqrt{2}(1+x^2)}(1+\sqrt{1-q^2})^{3/2}, \quad 
t_H=\frac{\sqrt{2}(1+x^2)}{x}\frac{\sqrt{1-q^2}}{(1+\sqrt{1-q^2})^{3/2}} \quad 
~{\rm with}~ \quad 
x=\frac{r_H}{R}.~~{~~}~~
\end{equation}
In   Figure \ref{EMBR}
we display 
the curves 
$q(a_H)$, 
$t_H(q)$ 
for several values of the ratio $r_H/R$.

 \begin{small}



 \end{small}

\end{document}